\documentclass[graybox, secnum]{svmult}

\usepackage{mathptmx}       % selects Times Roman as basic font
\usepackage{helvet}         % selects Helvetica as sans-serif font
\usepackage{courier}        % selects Courier as typewriter font
\usepackage{type1cm}        % activate if the above 3 fonts are
\usepackage{amsmath}                            % not available on your system
\usepackage{makeidx}         % allows index generation
\usepackage{graphicx}        % standard LaTeX graphics tool
\usepackage{subfig}         %to include subfigures
\usepackage{multicol}        % used for the two-column index
\usepackage[bottom]{footmisc}% places footnotes at page bottom

\usepackage{soul}            % for high-lighting of text
\usepackage{hyperref}        %for hyperlinks
\hypersetup{colorlinks=true,urlcolor=blue}

\usepackage{natbib}
\makeindex             % used for the subject index
\def\hi{{{\rm H}\,{\sc i}~}}

\def\civ{{{\rm C}\,{\sc iv}~}}

\def\cv{{{\rm C}\,{\sc v}~}}

\def\nvii{{{\rm N}\,{\sc vii}~}}

\def\oii{{{\rm O}\,{\sc ii}}}

\def\ovi{{{\rm O}\,{\sc vi}~}}
\def\ovin{{{\rm O}\,{\sc vi}}}
\def\ovii{{{\rm O}\,{\sc vii}~}}
\def\oviii{{{\rm O}\,{\sc viii}~}}

\def\neix{{{\rm Ne}\,{\sc ix}~}}

\def\oviin{{{\rm O}\,{\sc vii}}}
\def\oviiin{{{\rm O}\,{\sc viii}}}

\def\fexvii{{{\rm Fe}\,{\sc xvii}~}}

\def\xmm{{\it XMM-Newton}~}
\def\chandra{{\it Chandra}~}

\def\xmmn{{\it XMM-Newton}}
\def\chandran{{\it Chandra}}

\def\athena{{\it Athena}}
\def\arcus{{\it Arcus}}
\def\hirex{{\it HiReX}}
\def\xrism{{\it XRISM}}
\def\hubs{{\it HUBS}}
\def\lynx{{\it Lynx}}

\def\ka{{K$\alpha$}}
\def\kb{{K$\beta$}}
\def\lya{{Lyman-$\alpha$}}

\def\msun{{M$_{\odot}$}}

\def\kms{~km~s$^{-1}$~}
\def\gax{${_>\atop^{\sim}}$}
\def\lax{${_<\atop^{\sim}}$}
\def\gae{${_>\atop^{-}}$}

\def\1es{{1ES1553+113}}

\begin{document}
%\tableofcontents{}
\title*{Absorption studies of the most diffuse gas in the Large Scale Structure}
\author{Taotao Fang, Smita Mathur and Fabrizio Nicastro \thanks{corresponding author}}
\institute{Taotao Fang \at Department of Astronomy, Xiamen University, 422 South Siming Road, Xiamen, Fujian 361005, China\\ \email{fangt@xmu.edu.cn}
\and Smita Mathur \at The Ohio State University, 140 West 18th Avenue, Columbus, OH 43220, USA.\\ \email{mathur.17@osu.edu}
\and Fabrizio Nicastro \at Istituto	Nazionale	di	Astrofisica	(INAF)	–	Osservatorio	Astronomico	di	Roma,	via	Frascati 33,	Monte	  Porzio	Catone	00078,	RM,	Italy \\email:	fabrizio.nicastro@oa-roma.inaf.it \email{fabrizio.nicastro@oa-roma.inaf.it}}
\maketitle
\abstract{
As the Universe evolves, it develops a web of filamentary structure of matter. This cosmic web is filled with gas, with the most diffuse gas lying in the intergalactic regions. At low redshift, the gas is predominantly warm-hot, and one of its best tracers is X-ray absorption in sightlines to background quasars. In this Chapter, we present the theoretical background for the formation of the warm-hot intergalactic medium (WHIM) and present the physical properties of the WHIM from cosmological hydro-dynamical simulations. We discuss the feasibility of detecting the WHIM with X-ray absorption lines, with high-resolution and high signal-to-noise spectra. We present detailed discussion of observing techniques, including the WHIM ionization balance, observable lines, the curve of growth, and the diagnostics using the X-ray lines.  We present the current efforts of detecting the WHIM with gratings on-board \chandra and \xmm observatories. We discuss 
 the criticality of WHIM detections reported in literature, where robust detections are
 likely from the circumgalactic medium of intervening galaxies, or intra-group medium, rather than truly diffuse gas in the intergalactic medium. Secure detections of the most diffuse gas in the low redshift large scale structure may have to await next generation of X-ray telescopes. We end our Chapter with the discussion of future missions carrying dispersive and non-dispersive spectrometers. We present figure-of-merit parameters for line detectibility as well as for the number of WHIM systems that can be detected with future missions. These will define our ability to account for the missing low-redshift baryons and to understand the evolution of the Universe over half of its life. 
}

%\usepackage{}
%\section{Keywords} 
\keywords{X-ray absorption; IGM; WHIM; Large scale structure; Quasar absorption line systems}

\section{Introduction}\label{sec:intro}
%Introduction to the chapter; length depends on the topic describing importance of subject and content

In the past thirty years, the cosmological-constant ($\Lambda$)-dominated, cold dark matter ($\Lambda$CDM) model gradually became the standard paradigm in cosmology and astrophysics. In this model, the structure of the Universe grows linearly in the early epoch. In the later epoch, due to the influence of gravity, this process becomes non-linear, and determines the hierarchical structure formation process in which large structures are formed by merging smaller ones. This model has been successful in explaining a number of observations such as the cosmic microwave background (CMB) radiation and power spectrum of the mass distribution in the Universe.

A fundamental prediction of the $\Lambda$CDM model is the cosmic-web structure of the Universe at the largest scale: individual galaxies, galaxy groups, galaxy clusters, and gas in the intergalactic medium (IGM) are distributed in filamentary structures, interconnected with each other and separated by cosmic voids. This prediction has been at least partially confirmed by two observations: large-scale galaxy surveys such as the Sloan Digital Sky Survey (SDSS), and the Ly$\alpha$ forest observed in the spectra of high redshift quasars. Surveys of low-redshift galaxies show that indeed galaxies are distributed in the filamentary structures of the cosmic web (see, e.g., \citealp{davis82,york00}). The Ly$\alpha$ forest observations also reveal that at high redshift the majority of baryons are distributed in these filaments (\citealp{rauch97,weinberg97}).

However, significant problems exist to fully confirm this cosmic-web structure of the Universe, especially at low redshift. Surveys such as SDSS show the cosmic web distribution of galaxies; however, in terms of baryonic matter distribution, galaxies only represent the tip of the iceberg. Are there significant amounts of baryonic matter distributed in the filaments in-between galaxies, or in the outskirts of galaxies? If there are, why we haven't detected them yet? 

Coincidentally, detection of the diffuse gas in the large scale structure (LLS) may provide important clues to two significant ``missing baryons" problems, which refer to baryon shortfall at both LLS and galactic scales (e.g.,\citealp{fukugita98}, \citealp{mcgaugh10}, see \S\ref {sec:IGM}). Theory and numerical simulations suggest both ``missing baryons" may exist in the intergalactic gas, or in the outskirts of galaxies. Particularly, the interface between galaxies and the IGM, the so-called circumgalactic medium (CGM), may harbor a significant amount of baryons (e.g., \citealp{shull12,Wijers20,Gupta2012}). The CGM is discussed in details in Chapter 6, ``Probing the circumgalactic medium with X-ray absorption lines'', in Section XI of this Handbook. However, it is often difficult to distinguish gas distributed between galaxies or in the outskirts of galaxies. While we will primarily focus on the intergalactic gas, we will also briefly discuss the CGM gas whenever necessary.

Detecting the diffuse, intergalactic gas at low redshift is technically challenging. First, the most accessible line at the high-redshift Universe, the \hi Ly$\alpha$ line, shifts to the ultraviolet waveband and therefore is only observable with space telescopes. Secondly, the metal species that trace the cosmic web are difficult to detect due to their low number density and limited instrument capability. Cosmological  simulations show that the intergalactic gas at low redshift is heated to temperatures at $10^5$ -- $10^7$ K by shock waves from the structure formation process in and around galaxies. At this temperature, the most abundant element in the Universe, hydrogen, is nearly completely ionized and therefore undetectable (see \S\ref{sec:abs_obs}, Fig~\ref{f1_xtech}b, bottom panel). The only observables are the trace amounts of heavy elements such as carbon and oxygen (\S\ref{sec:abs_obs}, Figure~\ref{f1_xtech}b, top and middle panels) that are expelled from the stellar disks of galaxies in outflows. Typically, these heavy elements are highly ionized at this temperature, which makes them accessible in the Far ultraviolet (Far-UV) and X-ray wavebands only. 

In this chapter, we will review X-ray absorption studies of the most diffuse gas in the large scale structure (LLS). Similar to the Ly$\alpha$ forest in the optical band, highly ionized metal species leave imprints in the X-ray absorption spectra of background sources such as quasars. Since the absorption strength (e.g., the line equivalent width) is linearly proportional to the density of the ions (at least up to line saturation, see \S\ref{sec:cog}), absorption study is the most sensitive method to probe the extremely diffuse gas in the LLS, either in the IGM or in the CGM. 

In this chapter, we will not discuss absorption studies of the LLS in other wavebands, even though sometimes we will quote those studies for comparison. Research of the intergalactic gas in the nearby Universe became possible with Far-UV spectrometers on board the {\sl Hubble} Space Telescope and other space UV telescopes such as the Far Ultraviolet Spectroscopic Explorer ({\sl FUSE}). Many detected absorption lines such as \hi, \civ, \ovi reveal the rich content of the gas in the IGM and/or CGM. However, these ion species typically help probe the cold ($\leq 10^4$) K or warm-hot ($\leq 10^{5.7}$) K gas. Numerous results have been published in this area, and we refer readers to review articles such as \citet{McQuinn2016} for further reading. X-rays, on the other hand, primarily probe the hot ($\geq 10^{5.5}$) K content of the IGM and CGM, and will be the focus of this review chapter.

We will also not discuss X-ray emission studies of the diffuse gas in the LLS, for which we refer the reader to the chapter “Cluster outskirts and their connection to the cosmic web”,  within this section. Indeed, while X-ray emission can reveal the spatial distribution of the cosmic web structure, its emission is proportional to the square of density, and therefore it probes only the high density tail of the diffuse gas. Recent X-ray observations show that such high density gas is typically distributed in the outskirts of galaxy clusters (see, e.g., \citealp{vikhlinin03, eckert15, connor18, connor19}). The stacking technique in X-ray emission was sometimes applied to probe the very low density filaments (see, e.g., \citealp{tanimura20}). A number of other methods have been proposed to study the diffuse gas: for example, via the thermal Sunyaev-Zel'dovich effect (tSZ: i.e. interaction of the hot ionised gas in filaments with the CMB photons via inverse Compton scattering; e.g., \citealp{degraaff19}, \citealp{tanimura20}), or via the fast radio burst (see, e.g., \citealp{mcquinn14, prochaska19a, macquart20, lee21}). 

This chapter is organized as follows. In section \S\ref{sec:theory} we will provide a review of the main theory behind the hot, diffuse gas in the LLS and the history of X-ray absorption studies. Section \S\ref{sec:tech} will focus on the main techniques applied in X-ray absorption studies. In section \S\ref{sec:obs} we will review the main observational results of the past twenty years. Finally, we conclude in section \S\ref{sec:future} with the perspectives of several future X-ray missions.

\section{{Theory}} \label{sec:theory}
%chronologically from IGM to CGM \\

\subsection{History: A Hot Intergalactic Medium} \label{sec:history}

%The need for a significant among of IGM (and presumably hot): (1) To close the universe ($\Omega=1$); (2) To explain the lack of the Gunn-Peterson trough in quasar spectra; (3) To not violate the cosmic X-ray background.

The idea of a hot ($T\sim 10^6 K$), diffuse, intergalactic medium (IGM) was suggested in the 1960s and 1970s (see, e.g., \citealp{field72}) to help close the Universe (i.e., $\Omega_b \approx 1$), and to explain the lack of the suppression of the continuum in the blue side of the quasar Ly$\alpha$ emission lines (the Gunn-Perterson trough, \citealp{gunn65}). The neutral hydrogen atoms in the IGM would have produced such suppression, and the lack of it suggests hydrogen atoms are ionized due to the high temperature of the IGM. Later, it was proposed that, while most hydrogen and helium atoms are ionized in such a hot medium and cannot produce spectral features, highly ionized metals produced via pre-galactic nucleosynthesis or later star-formation processes, likely exist and can produce absorption lines and edges in the X-ray spectra of bright background X-ray sources, the so-called ``X-ray forest" (see, e.g., \citealp{sherman79, shapiro80, aldcroft94, hellsten98, perna98,fang00, fang02, chen03}), similar to the Ly$\alpha$ forest detected in the optical band. These early models, largely assuming a uniformly hot IGM, have been rejected by later observations since we now know that $\Omega_b \ll 1$, and the IGM only becomes hot in the local Universe. However, it is this development that motivates the later search of the X-ray absorption features in the LLS with high resolution, X-ray telescopes such as \chandra and \xmmn.

%The lack of such absorption trough can put constraints on the content of the baryonic matter in the Universe, and was applied to constrain cosmology models via the so-called ``X-ray Gunn-Perterson test" (e.g., \citealp{aldcroft94}).

%The detection of the soft X-ray background also suggested the IGM is likely to be hot to avoid absorption by the neutral hydrogen in the IGM (for a reviews see \citealp{field72}). 

%-----------------------------Figure Start------------------------------
\begin{figure}%
    \centering
    \includegraphics[trim=0 0 0 0,clip,width=8.0cm]{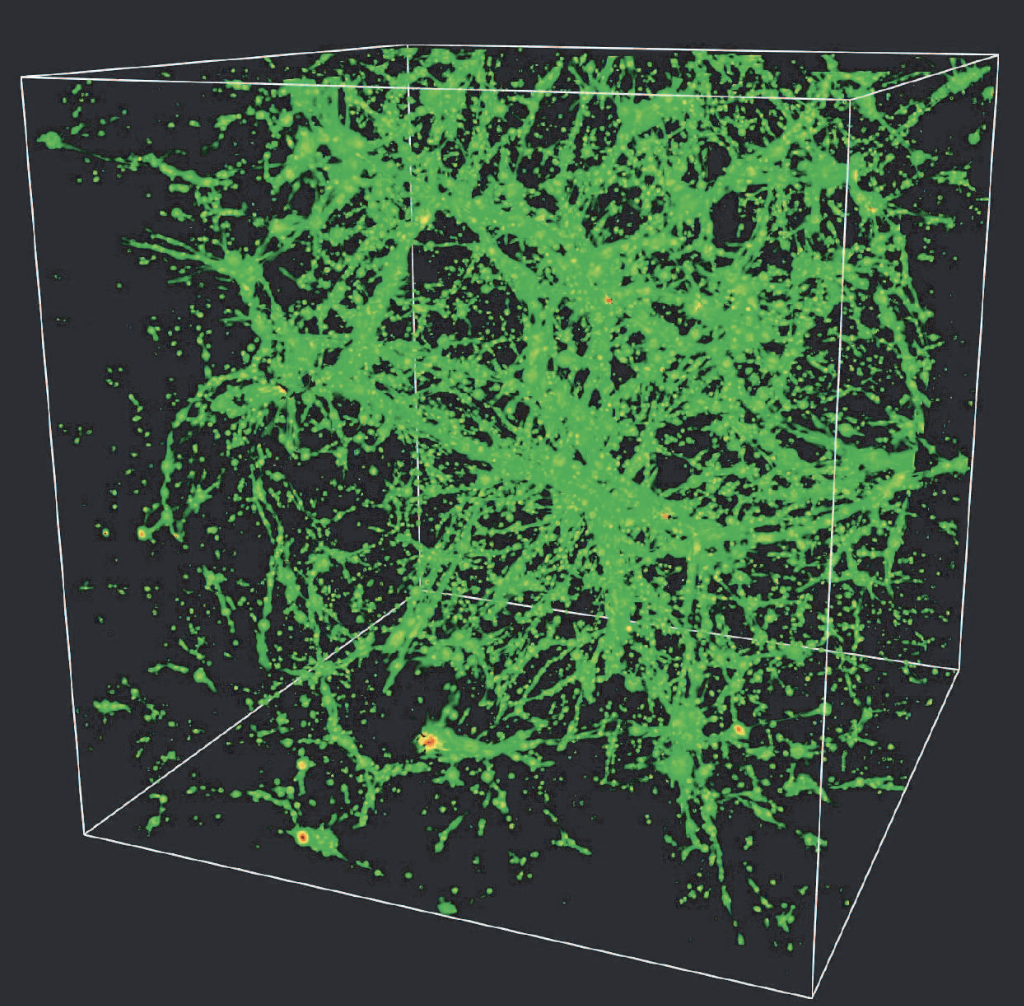}%
    %\vskip-2cm
    \caption{Cosmic-web structure of the Universe predicted by large-scale, cosmological hydro-dynamical simulations \citep{cen06a}.}%
    \label{fig:CenCosmicWeb}%
\end{figure}

\begin{figure}%
    \centering
    \includegraphics[width=10cm]{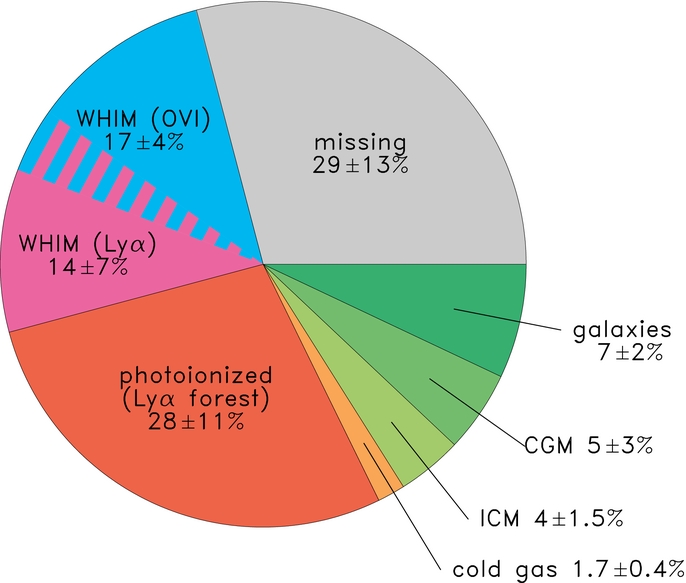}%
    \caption{Pie chart of the baryon distribution in the local Universe \citep{shull12}. Significant amount of baryons ($\sim 30-40\%$) are still missing.}%
    \label{ShullBaryonPieChart}%
\end{figure}

\begin{figure}%
    \centering
    \includegraphics[width=10cm]{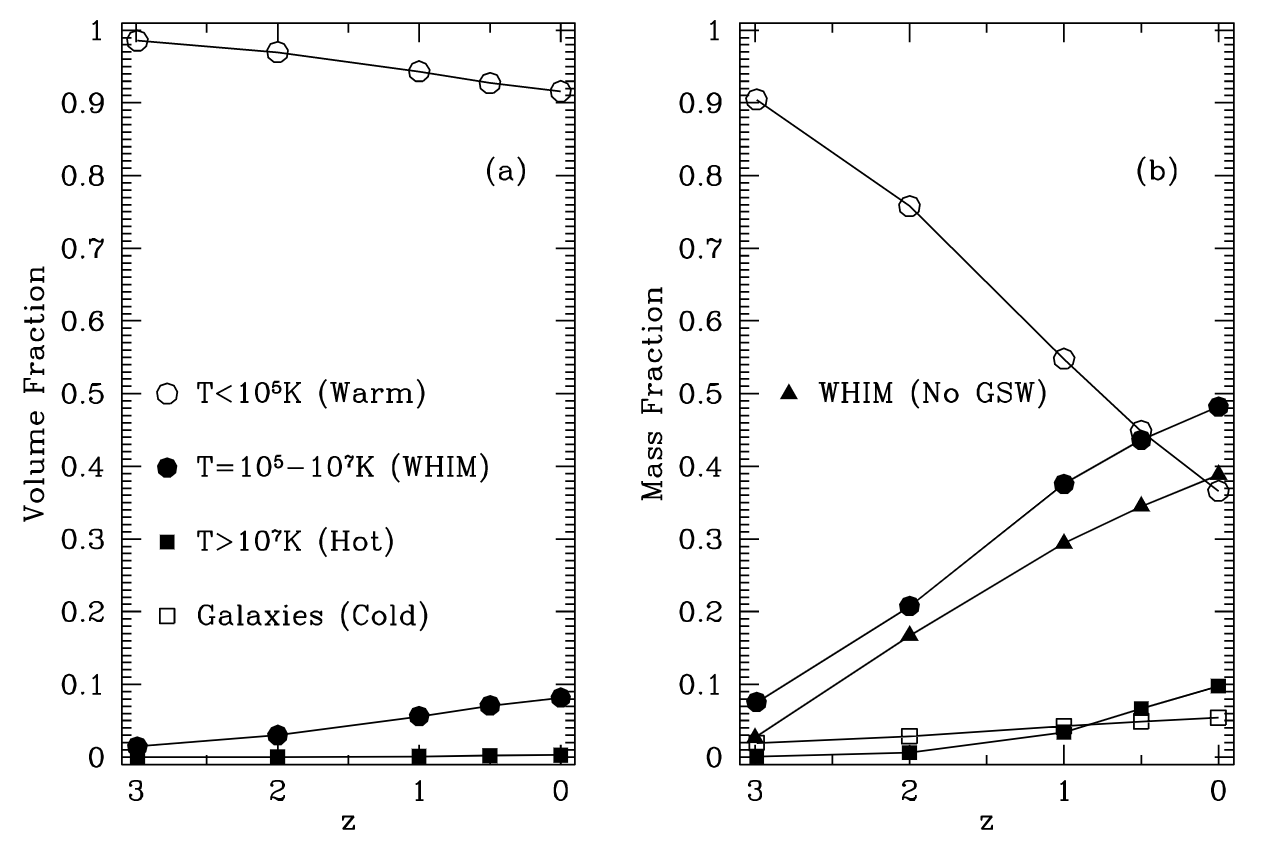}%
    \caption{Redshift evolution of baryons at different temperatures \citep{cen06a}. Left (right) panel shows the evolution of volume (mass) fractions. At high redshift ($z\sim3$), the vast majority of baryons ($>90$ \%) are in cool gas probed by the Ly$\alpha$ forest. However, in local Universe, baryons in warm-hot phase ($10^5 - 10^7$ K, filled circles in both panels) become dominant. Triangles in the right panels represent WHIM in a cosmological simulation without feedbacks from galactic super-winds (GSWs).}
    \label{CenBaryonEvolution}%
\end{figure}

\begin{figure}%
    \centering
    \includegraphics[width=10cm]{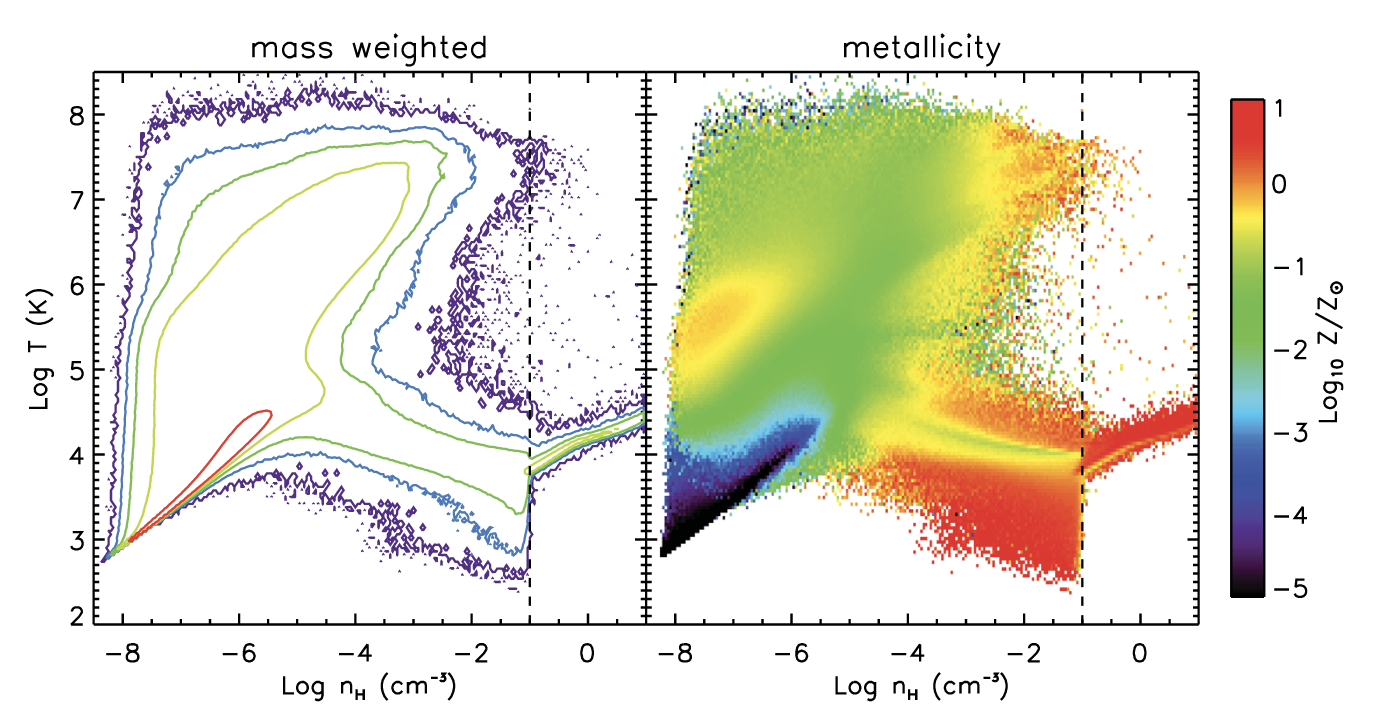}%
    \caption{Mass-weighted baryon phase diagram (left panel), and metallicity distribution (right panel) from hydro-dynamical simulations \citep{Bertone2013}.}%
    \label{BertoneMetallicity}%
\end{figure}

\begin{figure}%
    \centering
    \includegraphics[width=10cm]{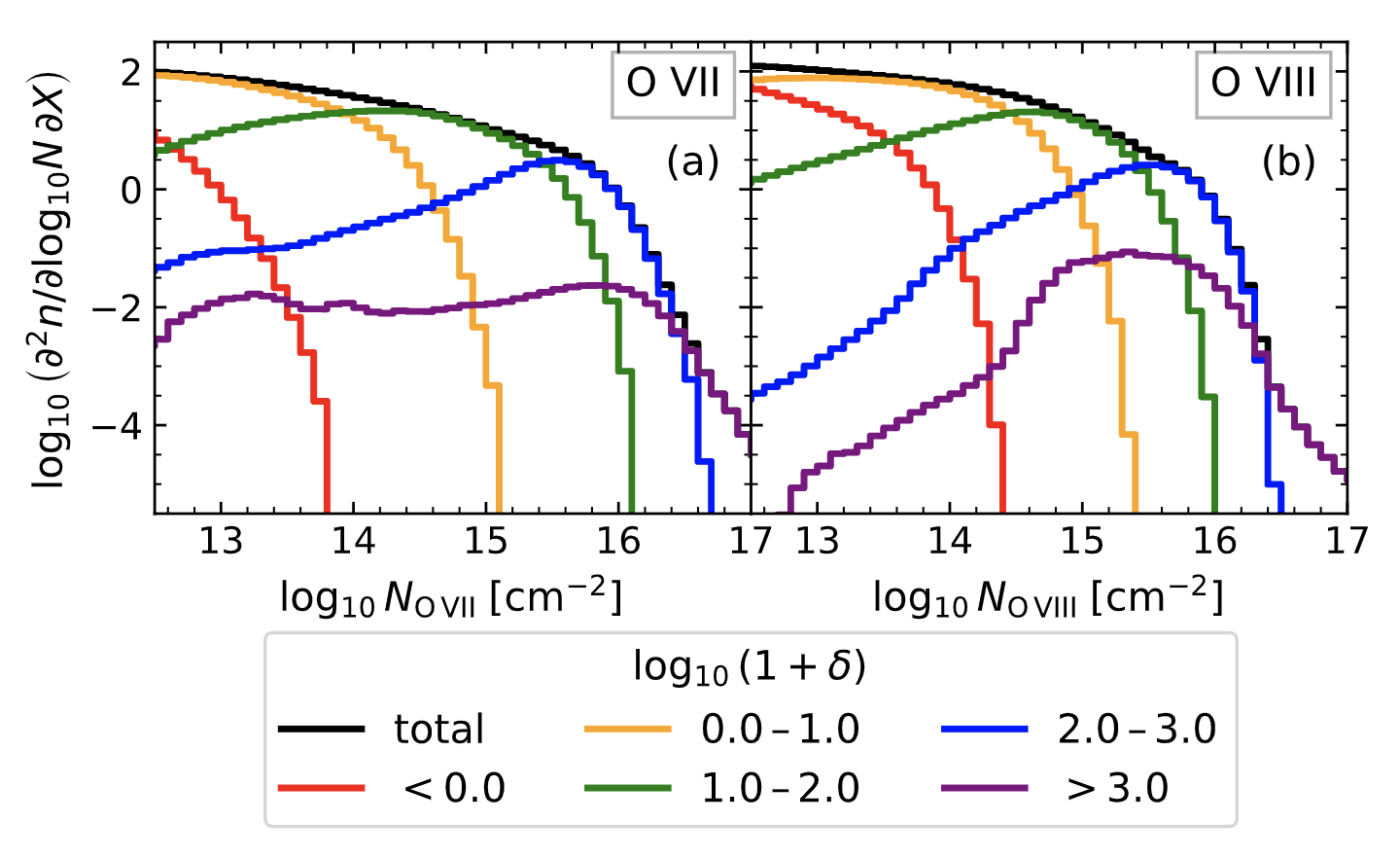}%
    \caption{Column Density Distribution Functions (CDDFs) of \ovii\ (left panel) and \oviii\ (right panel) absorption line systems in simulations \citep{Wijers2019}. Lines with the highest column density are produced by absorption systems with the highest overdensities.}%
    \label{WijersCDDF}%
\end{figure}

\begin{figure}%
    \centering
    \includegraphics[width=6cm]{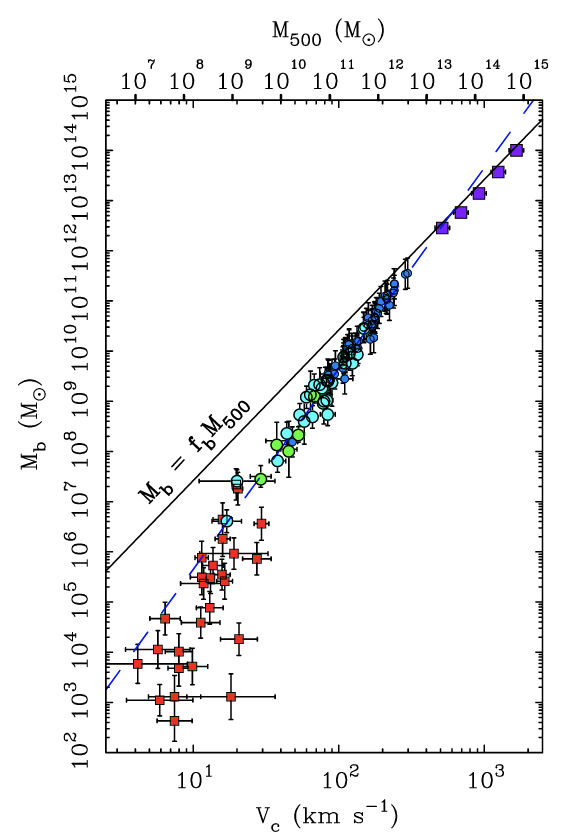}%
    \caption{Baryonic mass vs. rotation velocity for systems from dwarfs galaxies to galaxy clusters \citep{mcgaugh10}. The solid line shows the baryonic mass of a system if it followed the cosmic universal baryon fraction (17\%) of the dark matter mass.}%
    \label{McGaughMissingGalacticBaryon}%
\end{figure}
%-----------------------------Figure End--------------------------------

\subsection{The Large Scale Structure and the Warm-Hot Intergalactic Medium} \label{sec:IGM}

In the current most-accepted cosmology model, structures of the Universe form hierarchically via gravitational collapse, with small structures forming earlier. This ``concordance" model has been successfully proven by numerous observations in the past thirty years. Cosmological N-body plus hydro-dynamical simulations, based on this model, also predicted that while a fraction of baryons accretes to form collapsed structures, the majority of the matter in the Universe lies in the filaments and nodes of the cosmic-web form, and a significant amount may still reside in the intergalactic space (e.g., \citealp{Cen1999} and Figure~\ref{fig:CenCosmicWeb}).

Interestingly, the results from these numerical simulations provided a possible solution to the long-standing ``missing baryons" problem. In the 1990s, a number of studies suggested a shortage of unaccounted baryons in the low-redshift Universe (e.g., \citealp{Persic92, Bristow94, fukugita98}), by simply comparing the census of observed baryons in the local Universe with that indirectly derived from observations of the early Universe, i.e. the comparison of primordial deuterium abundance with big bang nucleosynthesis (BBN) models and the fluctuations of the CMB radiation.

Both, observations of the primordial deuterium abundance \citep{Cooke18} and the latest CMB measurements \citep{Planck20}, give a roughly consistent value of $\Omega_b \approx 0.045$, assuming a {\sl Hubble} constant of $70\rm\ km\ s^{-1}Mpc^{-1}$. 
However, in the local Universe, counting all the baryons in the collapsed phases (galaxies, stars) and in the photoionized IGM (Ly$\alpha$ absorbers), one can detect at most 60\% of the cosmic baryons predicted by BBN and CMB (\citealp{Persic92, Bristow94, fukugita98, shull12}), leaving a shortfall of $\sim 40\%$, the so-called ``missing baryons" (Figure~\ref{ShullBaryonPieChart}).

Cosmological hydro-dynamical simulations provided the first hint of where to find these missing baryons (\citealp{hellsten98,Cen1999,dave01,Fang2002, chen03, cen06a, CenFang2006, Branchini2009, Wijers2019}). Figure~\ref{CenBaryonEvolution} shows the volume (left panel) and mass (right panel) evolution of baryons as a function of redshift, for baryons at different temperature. At high redshift, the vast majority of the baryons (more than 90\%) are located in the IGM in the form of warm-cool ($T<10^5\ K$), diffuse gas. However, at low redshift, the gas in the temperature range of $10^5 - 10^7\ K$ dominates with a mass fraction of 40 -- 50\%. This diffuse gas ($n_b=10^{-6}-10^{-4}$ cm$^{-3}$ or over-densities $\delta = n/<n> = n/[2 \times 10^{-7}  (1+z)^3 (\Omega_bh^2/0.02)] = 5-500$) 
\footnote{We assume a Hubble constant of $H_0 = 100h\rm\ km\ s^{-1}Mpc^{-1}$, where $h=0.7$ throughout the chapter.}
permeates the intergalactic space in the LLS and forms the so-called ``warm-hot intergalactic medium," or WHIM \citep{Cen1999, dave01}. 

%The WHIM may account for 50-60\% of the gas in the IGM, but as much as $\sim 30$\% of the gas pervading the LSS at redshift $z$\lax{1} may be the residual photo-ionized \lya{ Forest} \citep{shull12}. 

It is suggested that while initially the gas in the high-redshift IGM remains mainly photo-ionized by the meta-galactic radiation field at temperatures below $10^5\ K$, the subsequent collapse of material towards dark-mater potential wells that feeds the formation of structures, triggers shock waves along the filaments, and heats the gas in the IGM to  temperatures in the range $10^5 - 10^7\ K$. Here we present a brief overview of theoretical predictions concerning the chemical and physical state of the WHIM and the potentials of  X-ray absorption as a probe of its existence. A detailed discussion from the observational and detectability point of view, will be presented in section \S\ref{sec:tech}.

%Because the largest WHIM volume fraction is distributed in low-density ($n_e \approx 10^{-6}-10^{-5}\rm\ cm^{-3}$) filaments, it is extremely difficult to detect them in emission, although there were reported detections of X-ray emission from bridges connecting massive galaxy clusters (i.e., \citealp{finoguenov03, eckert15}), or via stacking techniques (i.e., \citealp{tanimura20}). Absorption in the spectra of background sources is a more efficient method to probe these very low density regions, as the absorber column density scales only linearly with density. 

%The coolest fraction of the WHIM ($T<~10^{5.5}$ K) can be probed by UV spectroscopy with ion species such as \ovi and broad Ly$\alpha$ absorption. This has been demonstrated by numerous UV absorption studies with UV spectrometers onboard space telescopes such {\sl FUSE} and {\sl HST} (for reviews see, e.g., \citealp{Tripp08, Prochaska19b}). Recent studies (e.g., \citealp{shull12}) suggest that the WHIM probed in the Far-UV can account for 30\% of the total baryons (Figure~\ref{ShullBaryonPieChart}). However, the rest of the missing baryons are hotter ($T>10^6\ K$), and can only be probed in the X-rays with highly ionized species such as H and He-like oxygen. From the perspective of theory and numerical simulation, we present a brief overview of the metallicity, ionization, and X-ray absorption probe of the WHIM gas here. Detailed discussion from the prospective of observation and detectability will be presented in section \S\ref{sec:tech}.

%(e.g., Richter, Fang, \& Bryan 2006). Recent studies suggested that 

\textbf{Metallicity:} Metals, defined as elements heavier than hydrogen and helium, are important tracers of the WHIM. We still do not know precisely the metallicity of the diffused IGM, especially at low redshift, due to the lack of data. Numerical simulations suggest it could range between $10^{-4}$-- $10^{-1} Z_\odot$, where $Z_\odot$ is the solar metallicity (see Figure~\ref{BertoneMetallicity} and \citealp{cen06a, Oppenheimer08, Bertone13, Schaye15}). Metals in these simulations were typically produced via stellar processes such as Type II supernovae (SNe), Type Ia SNe and stellar mass loss in asymptotic giant branch stars, and then were transported out into the intergalactic space via galactic outflows powered by bursts of star formation and/or nuclear activity (Active Galactic Nuclei: AGNs). An important quantity, by an observational point of view, is the column density distribution of these metals, defined as their total number per unit area along a given line of sight. By comparing the column density distribution of metal species such as \civ and \ovi, it is shown that the predictions from numerical simulations agree reasonably well with observations, although details may vary depending on factors such as the strengths of galactic winds and other sub-grid physics (e.g.,\citealp{cen06a, Schaye15}).

\textbf{Ionization:} we know that both particle-collisions (during thermal shocks) and photo-ionization by the metagalactic ultraviolet background (UVB), must contribute to the ionization of the low-redshift IGM, by stripping its atoms of their electrons. The resulting ionized metals are generally named ``H-like" (hydrogen-like), ``He-like" (helium-like), ``Li-like" (lithium-like), etc., to reflect the number of bound electrons they are left with (i.e. one, two, three, etc., respectively). In modeling the exact ionization state of the IGM (and thus in estimating the column density distribution of a specific ion), however, two important assumptions have not yet been vigorously tested due to the lack of data. First, to account for photo-ionization simulations assume a theoretical metagalactic ultraviolet backgroud (UVB) such as that proposed by \citet{Haardt12}; however, several studies suggested the Haardt \& Madau UVB may underestimate the Hydrogen photoionization rates due to a low escape fraction of Lyman continuum photons (e.g., \citealp{kollmerier14, Shull15}). Such uncertainty may impact the production of \ovii and \oviii at column densities lower than $\sim 10^{16}\rm\ cm^{-2}$ as they likely trace the low density gas that is  predominately photo-ionized \citep{Wijers20}. Second, when the recombination timescale is shorter than the ion-electron equilibration timescale, non-equilibrium ionization may impact the production of ions. However, due to the low density of the WHIM, the effect of non-equilibrium ionization is typically not significant (see, e.g., \citealp{Kang05, Yoshikawa06, CenFang2006, Oppenheimer16, Wijers20}).

\textbf{X-ray absorption:} A useful quantity to describe the distribution of the predicted X-ray absorption line strengths, is the ``X-ray Forest Distribution Function" (XFDF, \citealp{hellsten98, fang02}), or equivalently ``Column Density Distribution Function" (CDDF, \citealp{Wijers20}): 
\begin{equation}
f(N^i,z) = \frac{\partial^2 n}{\partial \log_{10} N^i \partial z}.
\end{equation}
 This function represents the number of absorption systems (for the ion species $i$) per unit log column density per unit redshift. Figure~\ref{WijersCDDF} \citep{Wijers20} shows the CDDF (or XFDF) distributions of \ovii (left panel) and \oviii (right panel). Colored curves represent contributions from gas in different density regions. Interestingly, a break point is present in both panels at around $10^{16}\rm\ cm^{-2}$, suggesting absorbers with higher column density may arise in the (rarer to blindly cross) high over-density, galactic halo regions \citep{Wijers20}.

\subsection{The Circumgalactic Medium} \label{sec:CGM}

As noted in \S \ref{sec:intro}, a second ``missing baryons" problem exists on galactic scale: The cold baryon fraction (cold gas + stars) in Milky-Way size galaxies is well-below the mean cosmic baryon fraction (see, e,g, \citealp{mcgaugh10}, and Figure~\ref{McGaughMissingGalacticBaryon}). This ``missing galactic baryons" problem is closely related to the ``over-cooling” problem in the theory of galaxy formation and evolution (see, e.g., \citealp{White78, Maller2004}). In both semi-analytic and hydro-dynamical treatments of galaxy formation, hot gas cools to fuel ongoing star formation in Milky-Way-size galaxies. Standard treatments of cooling predict at least twice as much cold gas and stars in these galaxies as observed (see, e.g., \citealp{Benson03, Maller2004, Keres05}). Moreover, even in these models with “over-cooling,” most of the baryons in galaxy halos larger than $10^{11.3} M_\odot$ remain in the hot phase.

%It is quickly realized that a second ``missing baryons" problem exists on galactic scale. This ``missing baryons” problem, or the ``missing galactic baryons" problem, refers to the fact that the cold baryon fraction (cold gas + stars) in Milky-Way size galaxies is well-below the cosmic baryon budget ($\sim$ 20\%), and a large fraction of these missing galactic baryons are likely associated with extended, hot gaseous halos (see, e,g, \citealp{mcgaugh10}, and Figure~\ref{McGaughMissingGalacticBaryon}). This problem is closely related to the ``over-cooling” problem in the theory of galaxy formation and evolution (see, e.g., \citealp{White78, Maller04}). In both semi-analytic and hydro-dynamical treatments of galaxy formation, hot gas cools to fuel ongoing star formation in Milky-Way-size galaxies. Standard treatments of cooling predict at least twice as much cold gas and stars in these galaxies as observed (see, e.g., \citealp{Benson03, Maller04, Keres05}). Moreover, even in these models with “over-cooling,” most of the baryons in galaxy halos larger than $10^{11.3} M_\odot$ remain in the hot phase.

Proposed solutions to the over-cooling problem often involve strong feedback from supernovae and AGN (e.g., \citealp{hernquist03}), yet the impact that these prescriptions have on the surrounding gas remain largely untested. Another possibility is that the cooling itself is less efficient because the hot halos are more extended than previously suggested (\citealp{Maller2004, Fukugita06, SommerLarsen06}). Regardless of the specific solution, a large fraction of the baryons that naively would have cooled onto galaxies, must either be seating around galaxies in quasi-stable extended hot halos, or have been blown out of galaxy halos all together. In principle, about half of the still missing baryons in the Universe (i.e. $\sim 15-20$\% of the total baryons: see Fig. 2) could lie in the halos of galaxies (\citealp{Fukugita06, SommerLarsen06}). When considered in this context, constraints on the hot gas distribution within galaxy halos provide an important test of galaxy formation theory and probes a fundamental issue in cosmology.

The theoretical development of this idea of extended hot halo is, not surprisingly, coincident with the rapid development of multi-wavelength observations of outskirts of galaxies at the interface between their stellar disks and the IGM. These observations clearly demonstrate that the circumgalactic medium, or CGM, may harbor substantial amount of baryons, and may play a vital roles in regulating galaxy formation and evolution (for reviews, see \citealp{Bregman2007, Putman2012, tumlinson17}). Motivated by the latest observations, a number of analytic models were developed to explain multi-wavelength data (i.e., \citealp{Maller2004, Faerman2017, Faerman2020}). These models typically assume that gas in the CGM is in hydrostatic equilibrium with the galaxy's gravitational potential, and has a certain form of equation of state. After obtaining gas density and temperature profiles, these models can reasonably predict quantities such as absorbing column densities from Li- to H-like oxygen that are consistent with observations of our Milky Way. The CGM is discussed in detail in the Chapter  ``Probing the circumgalactic medium with X-ray absorption lines'' of Section XI of this Handbook. 

%Regardless of feedback and/or accretion mechanisms, recent models aimed at explaining the morphology and global properties of galaxies have applied increasingly on the idea of an extended, quasi-stable halo that contains a substantial amount of the CGM (e.g., \citealp{Keres05, Dekel06}). In the standard picture, the CGM around massive galaxies may contain the shock-heated gas that settles into a density profile that approximately traces that of the background dark halo. However, feedback heating will necessarily distort the density distributions of halo gas, affecting X-ray emission signals and the distribution and extent of cooling gas clouds. Similarly, early feedback or pre-heating before the halo collapses can affect halo gas profiles and cold gas halos in a related manner (\citealp{Mo02, Dave08}). The CGM in the extended galactic halo provides a potentially powerful probe of galaxy formation, gas fueling, and the regulating processes that govern galaxy evolution.

%\textbf{TBD: CGM simulation, cosmology vs. single galaxy}

%\textbf{TBD: CGM simulation, external galaxies vs MW size galaxies, from inside-out\\}

%\textbf{TBD: CGM theoretic modeling\\}

\section{{X-ray Techniques}} \label{sec:tech}
% include a table with lines, oscillator strength, typical EWs \\
%- include detectability formula, etc. 
\subsection{{Ionization Balance of the LSS gas in the Local Universe}} \label{sec:ionbal}
%As noted in \S \ref{sec:theory}, the low-redshift Universe is filled with warm-hot ($T=10^5-10^7$ K) and diffuse gas ($n_b=10^{-6}-10^{-4}$ cm$^{-3}$ or over-densities $\delta = n/<n> = n/[2 \times 10^{-7}  (1+z)^3 (\Omega_bh^2/0.02)] = 5-500$). The WHIM may account for 50-60\% of the gas in the IGM, but as much as $\sim 30$\% of the gas pervading the LSS at redshift $z$\lax{1} may be the residual photo-ionized \lya{ Forest} \citep{shull12}. 

\begin{figure}%
    \centering
    \subfloat[\centering \lya{-Forest} ionization balance]{{\includegraphics[width=10cm]{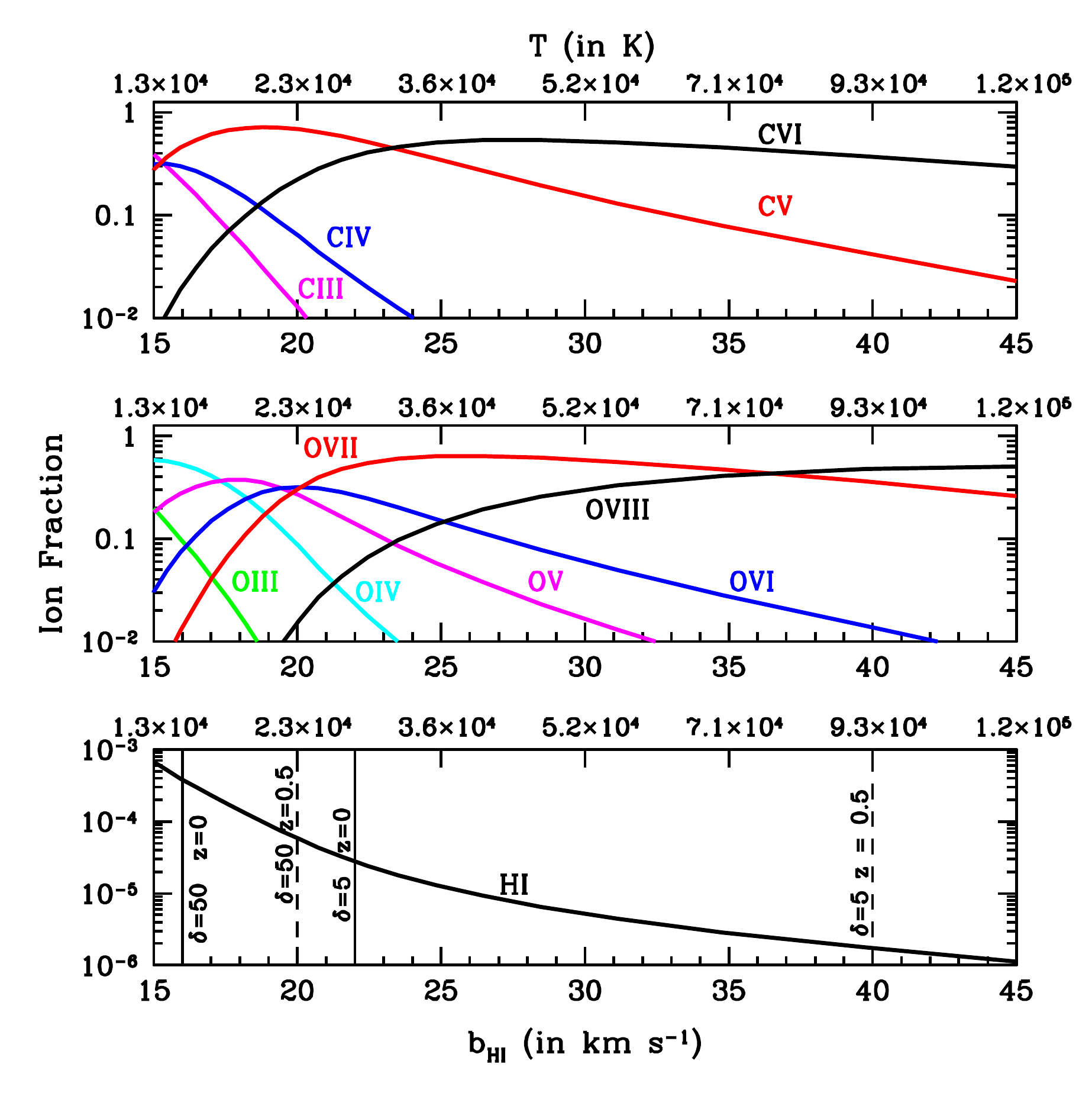} }}%
    \qquad
    \subfloat[\centering WHIM ionization balance]{{\includegraphics[width=10cm]{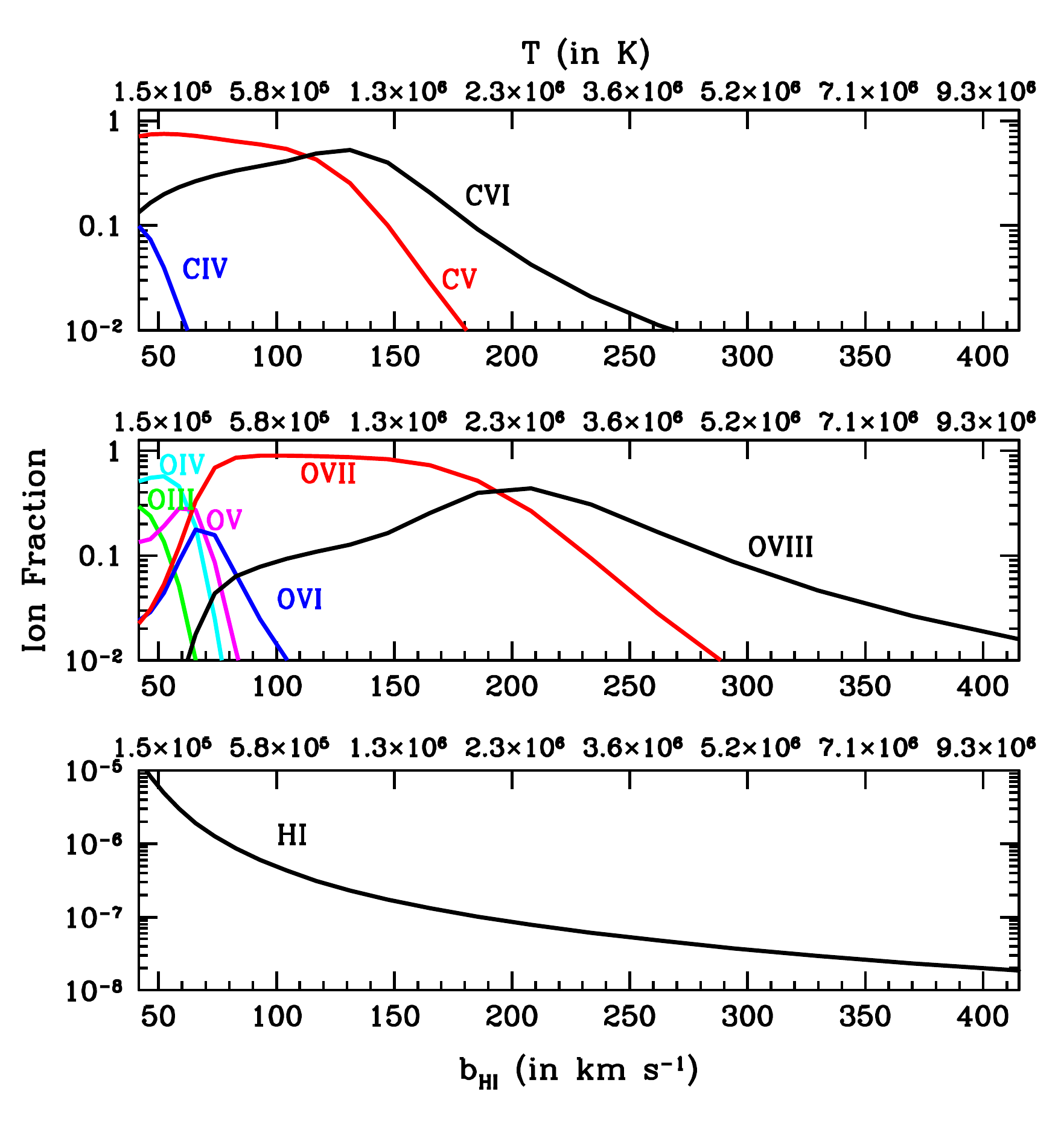} }}%
    \caption{Ionization balance of the most abundant ions of carbon (top panel), oxygen (central panel) and hydrogen (bottom panel), for (a) the \lya{ Forest} and (b) the WHIM, as a function of temperature (top x-axis) and HI Doppler parameter (bottom x-axis), i.e. the thermal width of hydrogen lines.}%
    \label{f1_xtech}%
\end{figure}

As noted in \S \ref{sec:theory}, the WHIM (including its, relatively dense, CGM fraction) may account for 50-60\% of the gas in the IGM, but as much as $\sim 30$\% of the gas pervading the LSS at redshift $z$\lax{1} may be the residual photo-ionized \lya{ Forest} \citep{shull12}. Both the gas producing the \lya{ Forest} and the WHIM contain highly ionized gas pervading  the space around and between the galaxies, but the mechanisms of ionization of these two important reservoirs of baryons and metals of the Universe are dramatically different. The gas in the \lya{ Forest} is purely photoionized by the meta-galactic radiation field (whose intensity depends on redshift), whereas the WHIM is mainly collisionally ionized through shocks during the collapse of density perturbation, but it also experiences the additional meta-galactic photoionization contribution, to a level that depends on its density. 
This yields to two different ionization structures: both phases contain highly ionized hydrogen and metals, but with different ion balances. 
Particularly, the \lya{ Forest} contains typically 1 neutral ion of hydrogen every $10^3-10^4$ atoms (Figure \ref{f1_xtech}a, bottom panel), and is therefore able to imprint strong \lya{} absorption in the far-ultraviolet (FUV) spectra of background quasars (e.g. \citeauthor{shull12}, \citeyear{shull12} and references therein), and the astrophysically most abundant metal (oxygen) is mainly between its C-like (i.e. with six bound electrons) and He-like (only two bound electrons) ions (Figure \ref{f1_xtech}a, central panel). 
The WHIM, instead, contains only 1 neutral atom of H every $10^6-10^7$ atoms (Figure \ref{f1_xtech}b, bottom panel), making it virtually transparent to FUV light, and oxygen is mainly distributed between its Li-like (three bound electrons) and fully-stripped ions (with the exception of a narrow range of temperatures at the lowest end of the WHIM temperature distribution, where C- through Be-like (four bound electrons) ions are relatively abundant: Figure \ref{f1_xtech}b, central panel). 

\begin{figure}%
\centering
\includegraphics[trim=10 10 0 10,clip,width=8.0cm]{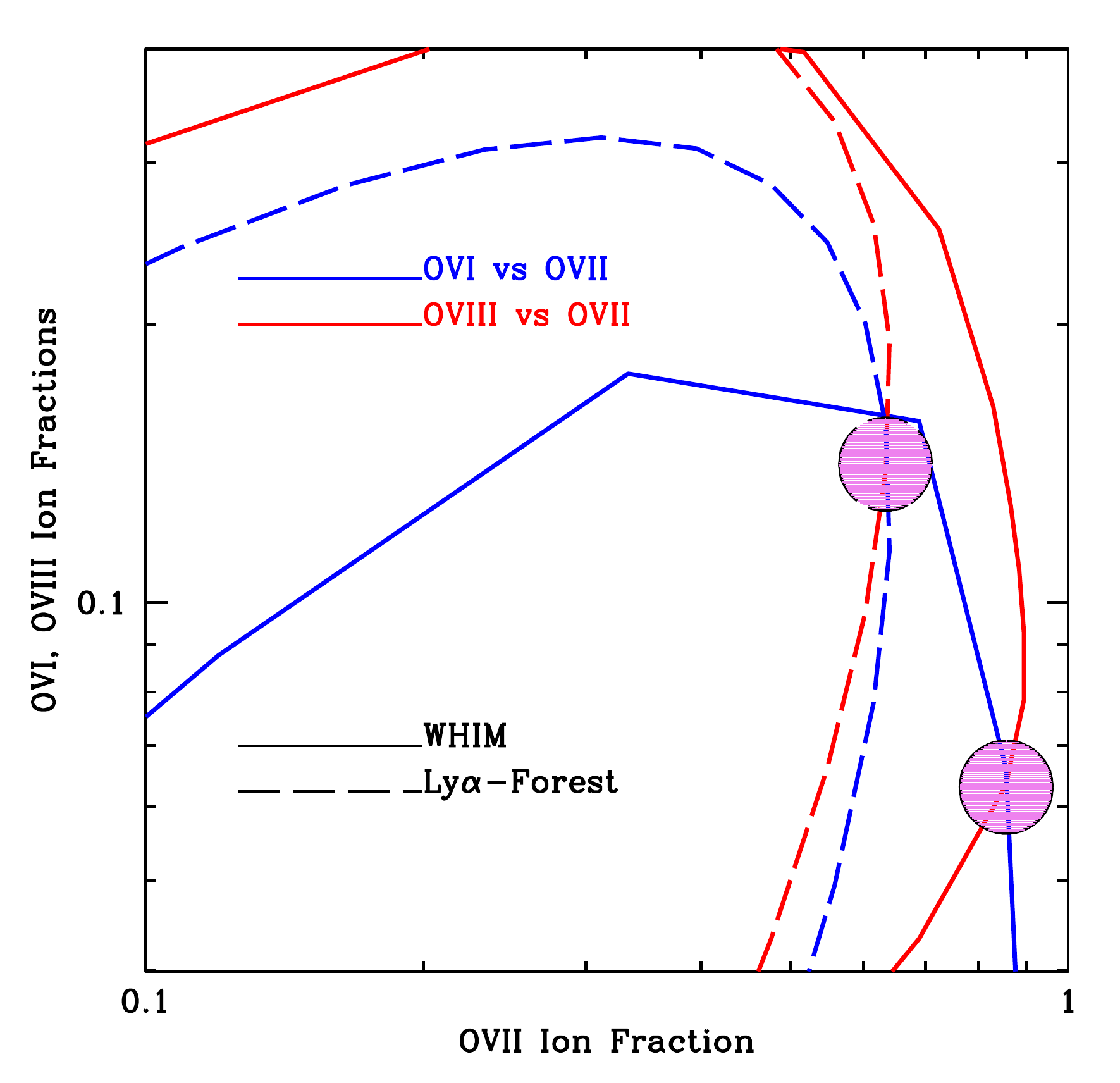}

\caption{fractions of Li-like (blue lines) and H-like (red lines) ions of oxygen as a function of the fraction of the He-like ion of the same element, for purely photoionized \lya{ Forest} gas (dashed lines) and hybridly-ionized WHIM with density of $n_e=10^{-5}$ cm$^{-3}$ (i.e. over-density $\delta=50$; solid lines). The two phases have dramatically different ionization balances, as highlighted by, e.g., the two pink circles, where same relative amount of Li- and H-like ions of oxygen correspond to $2.5\times$ different amounts of \ovii{}.}
\label{f2_xtech}
\end{figure}

\noindent
The different ionization structure of baryons and metals in the \lya{ Forest} and the WHIM is also shown in Figure \ref{f2_xtech}, where the fractions of Li-like (\ovi{}: blue) and H-like (\oviii{}: red) ions of oxygen are plotted as a function of the fraction of its stable He-like ion (\ovii{}). Dashed lines are for gas in the \lya{-Forest} conditions, while continuous lines refer to WHIM conditions. As an example, in Figure \ref{f2_xtech}, we highlight two regions (pink shaded circles) where each phase has Li- and H-like ions with similar fractional abundances but \ovii{} fractional abundances differing by a factor 2.5. 

%-----------------------------Figure Start------------------------------
%-----------------------------Figure End--------------------------------

%-----------------------------Figure Start------------------------------
%-----------------------------Figure End--------------------------------

These differences in the ionization structures of the two main reservoirs of baryons in the low-redshift Universe, makes it relatively easy to observationally distinguish between them, provided that we have adequate quality spectra. 
In the following we concentrate on the optimal detection techniques of gas in the WHIM, in the soft X-ray band, bearing in mind that most of the arguments we use for the WHIM,  and the same analysis techniques, apply to the low-$z$ photoionized \lya{-Forest} gas and that the two phases can be easily resolved through the above ionization-balance arguments. 

\subsection{{The WHIM Absorption Observables}} \label{sec:abs_obs}
Because of the temperature and density intervals that define the WHIM, the bulk of this matter (with the important exception of the low-temperature tail of its distribution) is virtually transparent to FUV light, where the outer-shell (i.e. excitation of an electron from the most outermost shell of an ion) resonant transitions (i.e. electronic transitions allowed in the atomic dipole approximation) of neutral hydrogen, neutral and once-ionized helium and several relatively low ionization state and light metals, lie. This is not the case for the soft X-ray radiation, which is opaque to metal electronic transitions from virtually all ionization stages.  

The soft X-ray band (namely E$\sim 0.2-3$ keV, or $\lambda \sim 4-120$ \AA) contains thousands of inner- (excitation of an electron from an inner-shell to a still occupied outer-shell) and outer-shell resonant transitions and ionization edges from all metal ions (neutral through H-like) from C through Fe. Table \ref{restran} lists the strongest of these resonant transitions from the most abundant ions at WHIM physical conditions. 

\begin{table}
%\scriptsize
\begin{center}
\caption{Strongest Soft X-ray (0.3-3 keV) Resonant Transitions at WHIM physical Conditions}
\label{restran}
\end{center}
\begin{tabular}{|llcccccc|}
\hline
 Element   &  Ion   & $^a$Abundance  & Transition & Rest Frame & Rest Frame & $^b$Oscillator Strength & $^c$Inner/Outer \\
    &              &  relative to H  & & Energy &  Wavelength & & \\
    &              &  (in $10^{-5}$) & & (in eV) & (in \AA) & & \\
\hline
%Hydrogen & H-like & 1 & $1s\rightarrow 2p$ (Ly$\alpha$)& $^a$10.199 & 1215.67 & $^a$0.416 & O \\
%Hydrogen & H-like & 1 & $1s\rightarrow 3p$ (Ly$\beta$)& $^a$12.088 & 1027.72 & $^a$0.079 & O \\
%Hydrogen & H-like & 1 & $1s\rightarrow 4p$ (Ly$\gamma$)& $^a$12.748 & 972.54 & $^a$0.029 & O \\
%\hline
%Helium & He-like & 0.0977 & $1s^2\rightarrow 1s2p$ (He$\alpha$)& $^a$21.218 & 584.33 & $^a$0.285 & O %\\
%Helium & H-like & 0.0977 & $1s\rightarrow 2p$ (Ly$\alpha$)& 40.814 & $^a$303.78 & $^a$0.416 & O \\
%\hline
Carbon  & Li-like & $24$ & $1s^22s\rightarrow 1s2s2p$ (K$\alpha_1$) & $^d$299.98 & 41.33 & $^d$0.483 & I \\
Carbon  & Li-like & $24$ & $1s^22s\rightarrow 1s2s2p$ (K$\alpha_2$) & $^d$303.44 & 40.86 & $^d$0.041 & I \\
Carbon  & Li-like & $24$ & $1s^22s\rightarrow 1s2s3p$ (K$\beta$) & $^d$336.50 & 36.85 & $^d$0.041 & I \\
Carbon  & He-like & $24$ & $1s^2\rightarrow 1s2p$ (He$\alpha$) & $^e$307.88 & 40.27 & $^e$0.648 & O \\
Carbon  & He-like & $24$ & $1s^2\rightarrow 1s3p$ (He$\beta$) & $^e$354.54 & 34.97 & $^e$0.141 & O \\
Carbon  & H-like & $24$ & $1s\rightarrow 2p$ (Ly$\alpha$) & $^e$367.47 & 33.74 & $^e$0.416 & O \\
Carbon  & H-like & $24$ & $1s\rightarrow 3p$ (Ly$\beta$) & $^e$435.49 & 28.47 & $^e$0.079 & O \\
\hline
Nitrogen  & Li-like & $7.95$ & $1s^22s\rightarrow 1s2s2p$ (K$\alpha_1$) & $^e$421.47 & 29.42 & $^f$0.431 & I \\
Nitrogen  & Li-like & $7.95$ & $1s^22s\rightarrow 1s2s2p$ (K$\alpha_2$) & $^f$425.45 & 29.21 & $^f$0.066 & I \\
Nitrogen  & He-like & $7.95$ & $1s^2\rightarrow 1s2p$ (He$\alpha$) & $^e$430.69 & 28.79 & $^e$0.675 & O \\
Nitrogen  & He-like & $7.95$ & $1s^2\rightarrow 1s3p$ (He$\beta$) & $^e$497.93 & 24.90 & $^e$0.144 & O \\
Nitrogen  & H-like & $7.95$ & $1s\rightarrow 2p$ (Ly$\alpha$) & $^e$500.34 & 24.78 & $^e$0.416 & O \\
Nitrogen  & H-like & $7.95$ & $1s\rightarrow 3p$ (Ly$\beta$) & $^e$592.93 & 20.91 & $^e$0.079 & O \\
\hline            
Oxygen  & Li-like & $49$ & $1s^22s\rightarrow 1s2s2p$ (K$\alpha_1$) & $^g$562.94 & 22.02 & $^g$0.328 & I \\
Oxygen  & Li-like & $49$ & $1s^22s\rightarrow 1s2s2p$ (K$\alpha_2$) & $^g$567.62 & 21.84 & $^g$0.067 & I \\
Oxygen  & He-like & $49$ & $1s^2\rightarrow 1s2p$ (He$\alpha$) & $^e$573.95 & 21.60 & $^e$0.696 & O \\
Oxygen  & He-like & $49$ & $1s^2\rightarrow 1s3p$ (He$\beta$) & $^e$665.55 & 24.90 & $^e$0.146 & O \\
Oxygen  & H-like & $49$ & $1s\rightarrow 2p$ (Ly$\alpha$) & $^e$653.62 & 18.97 & $^e$0.416 & O \\
Oxygen  & H-like & $49$ & $1s\rightarrow 3p$ (Ly$\beta$) & $^e$774.61 & 16.01 & $^e$0.079 & O \\
\hline
Neon  & He-like & $8.7$ & $1s^2\rightarrow 1s2p$ (He$\alpha$) & $^e$922.01 & 13.45 & $^e$0.724 & O \\
Neon  & He-like & $8.7$ & $1s^2\rightarrow 1s3p$ (He$\beta$) & $^e$1073.77 & 11.55 & $^e$0.149 & O \\
Neon  & H-like & $8.7$ & $1s\rightarrow 2p$ (Ly$\alpha$) & $^e$1019.33 & 12.13 & $^e$0.416 & O \\
Neon  & H-like & $8.7$ & $1s\rightarrow 3p$ (Ly$\beta$) & $^e$1210.91 & 10.24 & $^e$0.079 & O \\
\hline 
Magnesium  & He-like & $2.5$ & $1s^2\rightarrow 1s2p$ (He$\alpha$) & $^e$1352.24 & 9.17 & $^e$0.742 & O \\
Magnesium  & He-like & $2.5$ & $1s^2\rightarrow 1s3p$ (He$\beta$) & $^e$1579.31 & 7.85 & $^e$0.151 & O \\
Magnesium  & H-like & $2.5$ & $1s\rightarrow 2p$ (Ly$\alpha$) & $^e$1472.32 & 8.42 & $^e$0.416 & O \\
Magnesium  & H-like & $2.5$ & $1s\rightarrow 3p$ (Ly$\beta$) & $^e$1744.73 & 7.11 & $^e$0.079 & O \\
\hline 
Silicon  & He-like & $1.9$ & $1s^2\rightarrow 1s2p$ (He$\alpha$) & $^e$1864.98 & 6.65 & $^e$0.757 & O \\
Silicon  & He-like & $1.9$ & $1s^2\rightarrow 1s3p$ (He$\beta$) & $^e$2182.55 & 5.68 & $^e$0.152 & O \\
Silicon  & H-like & $1.9$ & $1s\rightarrow 2p$ (Ly$\alpha$) & $^e$1995.79 & 6.18 & $^e$0.416 & O \\
Silicon  & H-like & $1.9$ & $1s\rightarrow 3p$ (Ly$\beta$) & $^e$2376.45 & 5.22 & $^e$0.079 & O \\
\hline
Sulphur  & He-like & $1.2$ & $1s^2\rightarrow 1s2p$ (He$\alpha$) & $^e$2460.63 & 5.04 & $^e$0.767 & O \\
Sulphur  & He-like & $1.2$ & $1s^2\rightarrow 1s3p$ (He$\beta$) & $^e$2883.35 & 4.30 & $^e$0.153 & O \\
Sulphur  & H-like & $1.2$ & $1s\rightarrow 2p$ (Ly$\alpha$) & $^e$2621.23 & 4.73 & $^e$0.416 & O \\
Sulphur  & H-like & $1.2$ & $1s\rightarrow 3p$ (Ly$\beta$) & $^e$3107.37 & 3.99 & $^e$0.079 & O \\
\hline
Iron  & Ne-like & $2.7$ & $1s^22s^22p^6\rightarrow 1s^2s^22p^53d$ (L$\alpha$) & $^e$825.73 & 15.02 & $^e$1.95 & O \\
\hline
\end{tabular}
\noindent
a. \citet{Wilms2000}.\\ 
b. Absorption oscillator strength of the transition, defined by the relation $\mu_{lu} = A_{ul} \frac{g_u}{g_l} \frac{m_e c \lambda_{ul}^2}{8\pi^2 e^2}$, where $A_{ul}$ is the rate of spontaneous-emission from the upper ($u$) to the lower ($l$ levels of the transition (Einstein coefficient $A$), $g_l$ and $g_u$ are the statistical weights of the lower and upper levels of the transition, respectively, and $\lambda_{ul}$ is the rest-frame wavelength of the transition. \\
c. Inner- (I) or Outer-shell (O) transition (see text for details). \\
d. Experiment: \citet{Muller2009}. \\
e. Theory: \citet{Verner1996}. \\
f. Experiment: \citet{Shorman2013}. \\
g. Experiment: \citet{McLaughlin2017}. \\
\end{table}

\noindent
When the soft X-ray light from a background source (e.g. a quasar) intercepts chemically enriched gas along the line of sight, metals in the gas undergo resonant scattering by absorbing those line-of-sight photons with wavelengths $\lambda$ equal to their resonant transitions wavelengths, and re-emitting them randomly in all directions (on timescales given by the inverse of their Einstein coefficients for spontaneous emission, proportional to the transition oscillator strength: see, footnote $^b$ of Tab. \ref{restran} for  definitions of these two quantities and, e.g., \citealp{RybickiLightman85} for details). The net effect is to take away those photons from our line of sight, and thus produce an absorption line in the spectrum of the background source. 

\noindent
For a given ion $i$ of the chemical element $X$ undergoing a $l\rightarrow u$ electronic transition, the two quantities directly observable/measurable are the strength of the line relative to the continuum (or equivalent width, $EW$) and the absorption line profile $\phi(\lambda)$. The two quantities are, generally, not independent of each other. 

\begin{figure}%
    \centering
    \subfloat[\centering \ovii{He-$\alpha$} line profiles]{{\includegraphics[width=5cm]{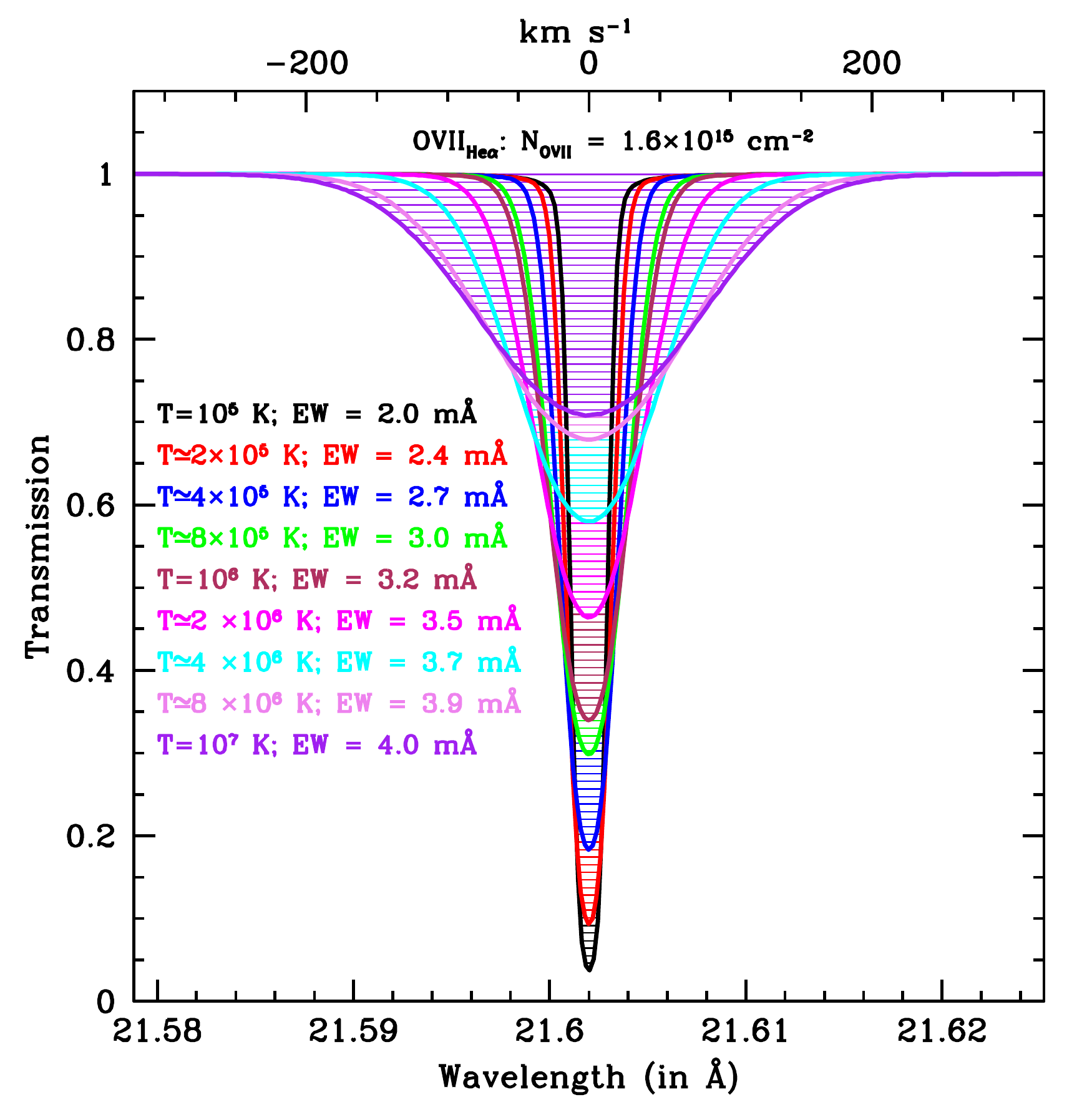} }}%
    \qquad
    \subfloat[\centering \ovii{He-$\beta$} line profiles]{{\includegraphics[width=5cm]{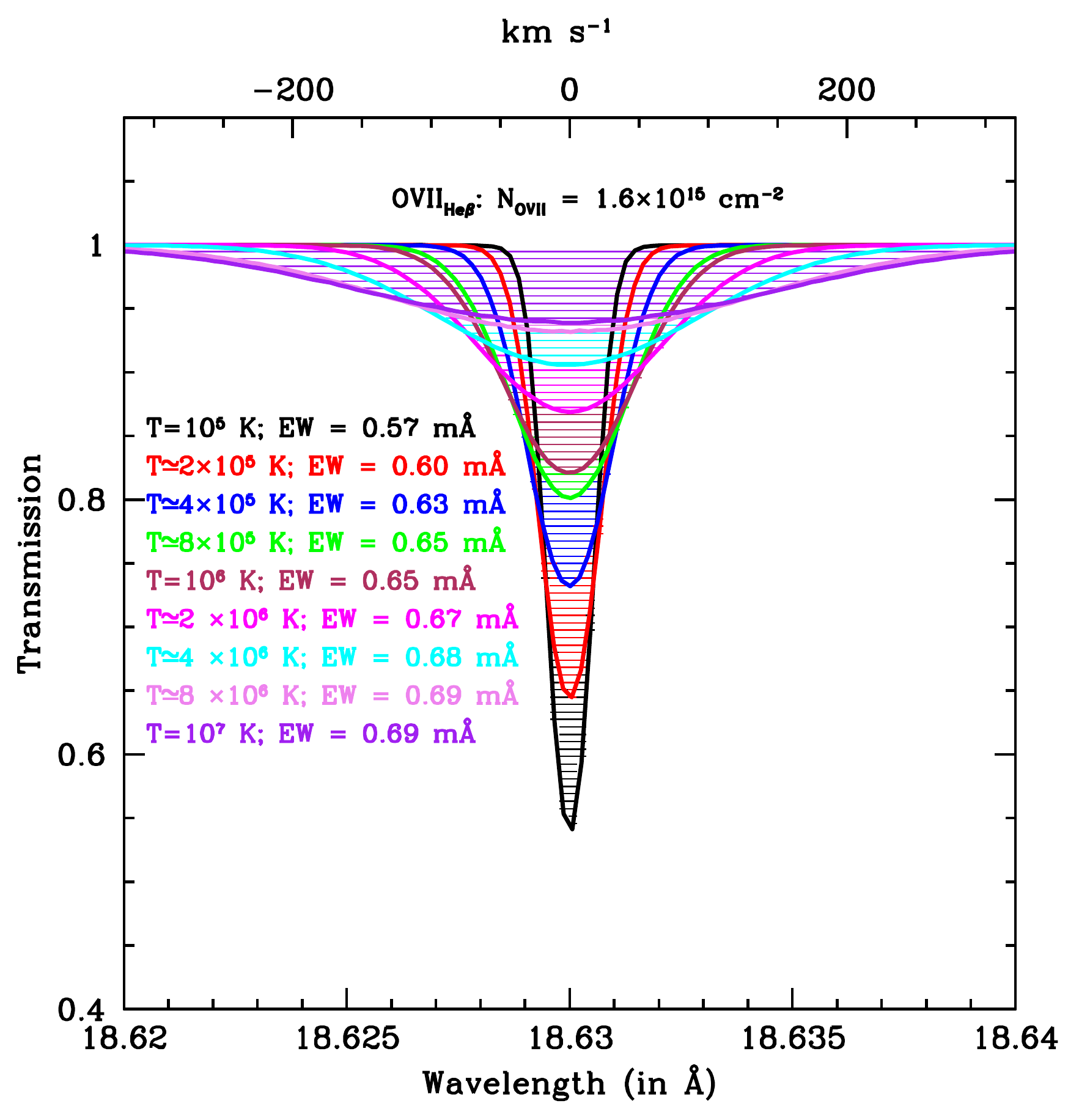} }}%
    \caption{Nine \ovii{$_{He-\alpha}$} (left panel (a)) and \ovii{$_{He-\beta}$} (right panel (b)) absorption line profiles imprinted by gas with \ovii{} column density N$_{OVII} = 1.6\times 10^{15}$ cm$^{-2}$, and temperatures ranging from T$=10^5-10^7$ K ($b_{O} = 10-100$ km s$^{-1}$). The non-thermal line broadening is assumed to be zero.
    }
    \label{f3_xtech}
\end{figure}

\noindent
The line $EW$ is given by the ratio between the integrated line flux (assumed to be positive for absorption lines: lined areas in Figure \ref{f3_xtech}a,b) and the monochromatic continuum flux at the line center normalized to 1 (see Figure \ref{f3_xtech}a,b).  

\noindent
The absorption line profile is a combination of Gaussian and Lorentzian. 
The Gaussian broadening is due to thermal and non-thermal motions in the gas. The thermal, Doppler parameter of a line imprinted by the ion $X^i$, is given by: 
\begin{equation} 
b_{X^i}^{th}(T) = (2kT/m_A)^{1/2} = (2kT/A m_p)^{1/2} = b_{HI}^{th}(T)/A^{1/2},
\end{equation} 
where A is the atomic weight of the ion, $m_p$ is the proton mass and $b_{HI}^{th}(T)$ is the thermal Doppler parameter for hydrogen. 
The non-thermal motions could be due to turbulence or line-of-sight differential (e.g. the Hubble flow, for line-of-sight extended filaments) motions $\Delta \textrm{v}$. These two mechanisms combine together in quadrature to provide an effective Doppler parameter: 
\begin{equation}
b = \sqrt{[b_{X^i}^{th}(T)]^2 + (\Delta \textrm{v})^2}. 
\end{equation}

The Lorentzian broadening $\gamma$ is an intrinsic atomic feature of the given electronic transition and is implied by the uncertainty principle, due to the finite (non-zero) duration of an excited electronic state. The natural width of a transition $\gamma$ is proportional to the rate of decay of the upper level of the transition (into all possible lower-energy channels), i.e. to the oscillator  strength $\mu_{lu}$ of the transition.  
These two distributions combine to provide a so called 'Voigt' line profile (i.e. the average of the intrinsic Lorentzian line profile weighted by the Maxwellian particle velocity distribution: see e.g. \citeauthor{RybickiLightman85}, \citeyear{RybickiLightman85}, for a full treatise).  
The ratio between the Lorentzian and Gaussian widths, defined as parameter $a$, determines the degree of saturation of a given line, and thus relates the two observables (line profile and Equivalent Widths). 

Figure \ref{f3_xtech}a,b show nine \ovii{$_{He-\alpha}$} and \ovii{$_{He-\beta}$} absorption line profiles imprinted by an \ovii{} column density N$_{OVII} = 1.6\times 10^{15}$ cm$^{-2}$, in gas with temperatures ranging from T$=10^5-10^7$ K (oxygen thermal Doppler parameter $b_{O} = 10-100$ km s$^{-1}$) and absence of turbulence or line-of-sight differential motion ($\Delta \textrm{v}=0$). The corresponding $a$ parameters of these two transitions range in the intervals $0.0056-0.056$ (Figure \ref{f3_xtech}a) and $0.0014-0.014$ (Figure \ref{f3_xtech}b), respectively. The He-$\alpha$ transition is intrinsically 4.8 times stronger than the He-$\beta$ (Table \ref{restran}). This translates into shallower and broader (Gaussian dominated) He-$\beta$ lines, compared to He-$\alpha$. 

%-----------------------------Figure Start------------------------------
%-----------------------------Figure End--------------------------------

The absorption line profiles of Figure \ref{f3_xtech}a,b are displayed at a spectral resolution of 0.2 m\AA\ (resolving powers $R=\lambda / \Delta\lambda \simeq 100,000$ at soft-X-ray  wavelengths, corresponding to velocity resolutions of $\sim 3$ km s$^{-1}$, oversampling WHIM oxygen-line widths by factors \gae{15}) and infinite signal-to-noise per resolution element (SNRE). In these ideal conditions, line profiles and $EW$s can be directly and independently measured and thus ion column densities directly derived from single-line measurements. 

Observationally, however, the situation is more complicated. Not only infinite SNRE is impossible to be attained, but the resolutions of the current (and future) generation of X-ray spectrometers are at least 100 times worse than those used in Figure \ref{f3_xtech} (see Table \ref{tab:future_spec} in \S \ref{sec:future}). The intrinsic degeneracy between line profiles and $EW$s is hidden into unresolved absorption line spectra, but can be removed by resorting to curve-of-growth techniques, applied to at least two different transitions from the same ion. 

\subsection{Absorption Line Curves of Growth} \label{sec:cog}
\begin{figure}%
\centering
\includegraphics[trim=10 10 0 10,clip,width=10.0cm]{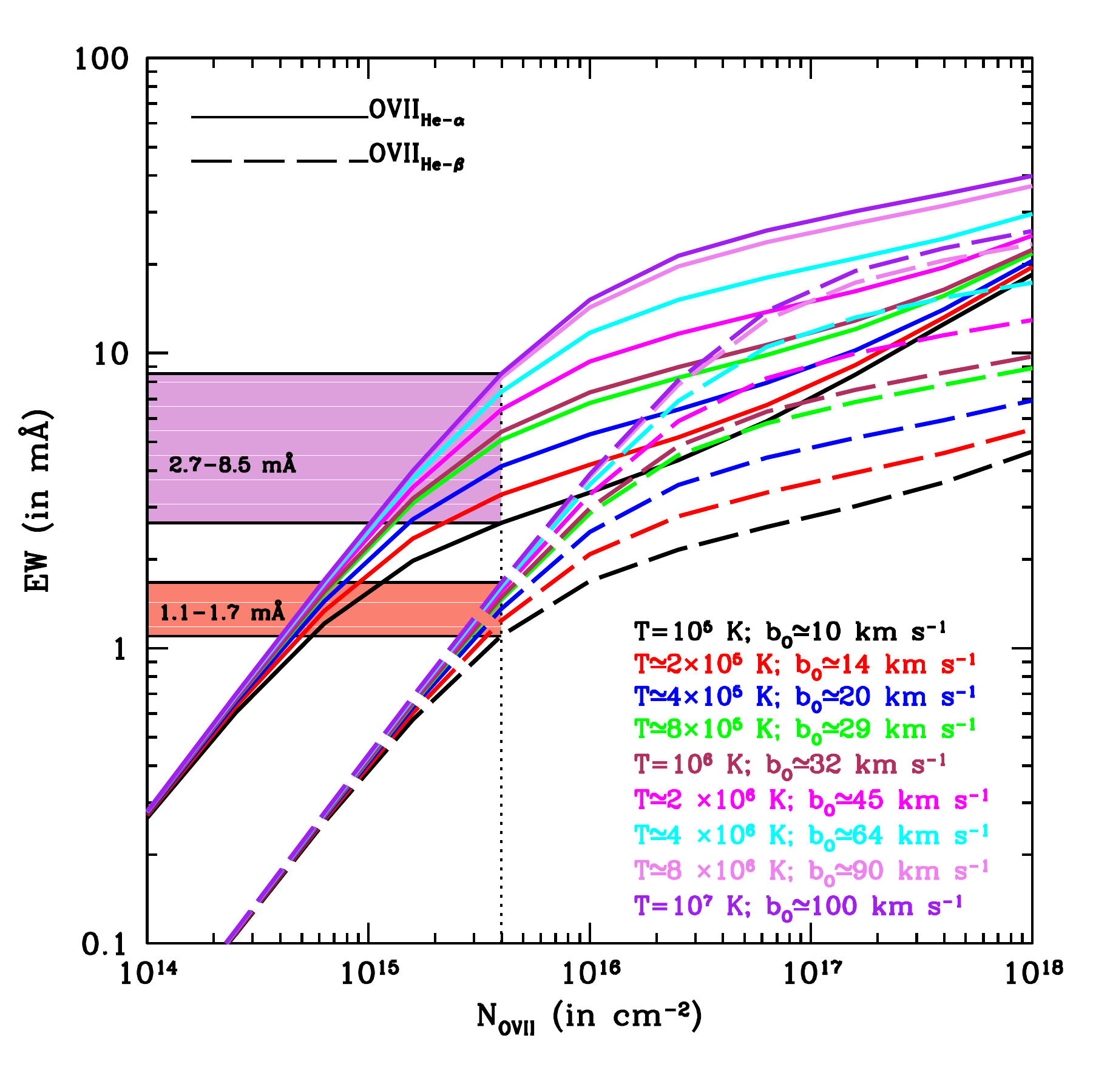}

\caption{Curves of growth of the two transitions of Figure \ref{f3_xtech}a,b (solid and dashed curves, respectively), for oxygen Doppler widths in the range $b_{O} = 10-100$ km s$^{-1}$ (as labeled) and no turbulence or line-of-sight differential motion (i.e. $\Delta \textrm{v} = 0$). The two shaded areas show the ranges of allowed $EW$s of the \ovii{} $He-\alpha$ (violet area) and $He-\beta$ (orange area) transitions, corresponding to an \ovii{} column density N$_{OVII} = 4\times 10^{15}$ cm$^{-2}$.} 
\label{f4_xtech}
\end{figure}

The curve of growth of an absorption line is the growth of line $EW$ with increasing ion column density and, as shown in Figure \ref{f3_xtech}a,b, depends on the intrinsic strength of the transition and the physics and kinematics of the absorbing gas. 
Figures \ref{f4_xtech}a,b, show the curves of growth of the two transitions of Figure \ref{f3_xtech}a,b, respectively, for oxygen Doppler widths in the range $b_{O} = 10-100$ km s$^{-1}$ (and no turbulence or line-of-sight differential motion: $\Delta \textrm{v} = 0$). 

At low ion column densities (below a threshold that depends on the intrinsic strength of the transition, namely its oscillator strength), $EW$s grow linearly with the number of line-of-sight particles. In this portion of the curve of growth, line $EW$s are independent on the line profile, and can be analytically approximated by eq. \ref{eq:ew_lin}: 
\begin{equation}
    EW \simeq 8.85\times 10^{-18} \textrm{N}_{X^i} \lambda_0^2 \mu_{X^i} \textrm{ m}\AA, 
\label{eq:ew_lin}
\end{equation}
where N$_{X^i}$ is the column density of the ion $i$ of the element $X$ in cm$^{-2}$, $\lambda_0$ is the central wavelength of the transition in \AA\ and $\mu_{X^i}$ its oscillator strength. 
%and $z$ the redshift of the absorber. 
\noindent
Above the column density threshold, absorption lines start saturating and their curves of growth bend into a non-linear regime, whose precise path in the $EW$-N$_{X^i}$ plane depends critically on the line profile (and thus width). 

\noindent
The non-linear growth of absorption line $EW$s continues till a second (much higher and visible in Figure \ref{f4_xtech} only for the lowest temperature/thermal-broadening curves of growth) column density threshold, above which lines fully saturate to the zero-flux level and their $EW$s start increasing again with the square root of the ion column density (see \citep{Spitzer1978} for details). 

\begin{figure}%
\centering
\includegraphics[trim=10 10 0 10,clip,width=10.0cm]{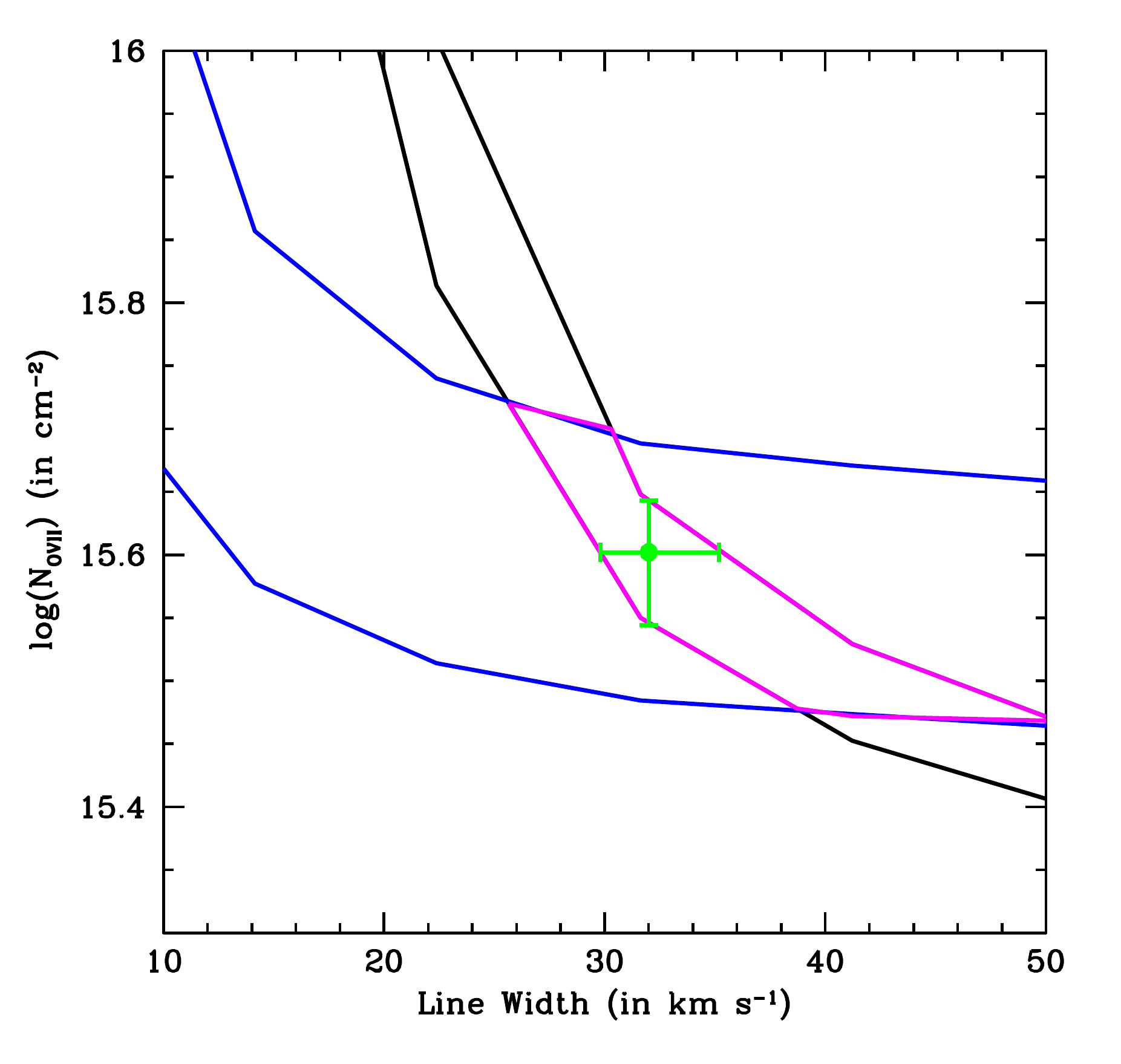}

\caption{90\% \ovii{}$_{He-\alpha}$ (black lines) and \ovii{}$_{He-\alpha}$ (blue lines) $b-log(N_{\rm OVII})$ contours for measured $EW(He-\alpha) = 5.4 \pm 0.3$ m\AA\ and $EW(He-\beta) = 1.5 \pm 0.3$ m\AA. The green point and error bars show the best solution, while the magenta contours show the  90\% solution for 2-interesting parameters.}
\label{f5_xtech}
\end{figure}

In the saturated regimes, the $b-$N$_{X^i}$ degeneracy can be removed by measuring the $EW$ of two transitions of the same ion. As an example, the shaded areas of Figure \ref{f4_xtech} show the range of allowed $EW$s of the \ovii{} $He-\alpha$ and $He-\beta$ transitions, corresponding to an \ovii{} column density N$_{OVII} = 4\times 10^{15}$ cm$^{-2}$. Determining, with sufficient accuracy, the $EW$s of the two transitions, allows one to find the unique $b-$N$_{X^i}$ solution (see, e.g. Figure \ref{f5_xtech}).  

%-----------------------------Figure Start------------------------------
%-----------------------------Figure End--------------------------------

%-----------------------------Figure Start------------------------------
%-----------------------------Figure End--------------------------------

\subsection{WHIM Gas Diagnostics} \label{sec:gas_diag}
With ion column densities and line widths in hands, the gas physical (e.g. ionization balance, and thus temperature, density, mass), kinematic (e.g. bulk and turbulent motion) and chemical (relative abundances) conditions can be derived. 

\subsubsection{WHIM Physical Conditions} \label{sec:whim_ionbal}
The column density of the ion $i$ of the element $X$ is given by the product between the baryon equivalent column density N$_b$, the fractional abundance of the ion $i$ relatively to all ions of its element, $f_{X^i}$,  and the chemical abundance of the element $A(X)$: 

\begin{equation} \label{eq:ioncol} 
\textrm{N}_{X^i} = \textrm{N}_b f_{X^i} A(X). 
\end{equation}

Thus, the ratio of different ion column densities from different ions of the same element allow us to determine the gas ionization balance $\xi(f)$ = ($f_{X^i}/f_{X^j})$ (for $i\neq j$ and all elements $X$), and thus the physical state of the intervening gas and the mechanisms of ionization (pure collisions, pure photo-ionization or an hybrid combination of the two). 

In the case of photo- or hybrid-ionization the source of photo-ionization are photons from the meta-galactic radiation field, which is relatively well known at $z$\lax{2} (e.g. \citeauthor{shull99}, \citeyear{shull99}; \citeauthor{Haardt12}, \citeyear{Haardt12}; 
but see \S \ref{sec:IGM} for possible sources of uncertainties). Thus, by modeling the measured gas ionization balance $\xi(f)$ with proper photo- or hybrid-ionization models, the  density (or ionization parameter) and temperature of the gas can be estimated (see e.g. bottom panel of Figure \ref{f1_xtech}a). 
For purely collisionally ionized gas (over-densities $\delta$ \gax{500}), the temperature of the gas is uniquely determined by the measured fractional ion abundances distribution $\xi(f)$. 

\subsubsection{WHIM Kinematics} \label{sec:t_and_kin}
The bulk line-of-sight velocities of the absorbing gas (or gas components, where multiple phases co-exist in the same structure) relative to, e.g., the redshift of the nearest galaxy, can be directly estimated by measuring the absorption line centroids, down to precisions that depend both on the spectrometer resolution and the S/N ratio at the line negative peaks. Velocity differences between different components of the same structure will inform us on the relative motions between physically different phases, e.g. the motion of warm (or even cold) condensations embedded in hotter and less dense filaments. 

The widths of the absorption lines (or line components), instead, can in principle tell us about the ion kinetic temperature of the gas and line-of-sight turbulence or differential motions. %At very high X-ray spectral resolution, the temperature and line-of-sight kinematics of the absorbing gas can also directly be estimated by comparing the width of absorption lines imprinted by different elements. 
As shown in \S \ref{sec:cog} the width of the line $b$ contains two Gaussian broadening terms, caused by the thermal $b_{X^i}^{th}$ and turbulence (or line-of-sight differential) $\Delta \textrm{v}$ particle motion. For the generic ion $i$ of the elements $X$, we can write: 

\begin{equation} \label{eq:dop}
b_{X^i} = \sqrt{[b_{X^i}^{th}(T)]^2 + (\Delta \textrm{v})^2} = \sqrt{[2kT/(A(X)m_p)] + (\Delta \textrm{v})^2},  
\end{equation}

The thermal Doppler parameter depends on the atomic mass of the element imprinting the absorption line, while $\Delta \textrm{v}$ is the same for any ion. Thus with sufficiently high spectral resolution, by accurately measuring $b$ for at least two absorption lines from different elements $X$ and $Y$ (with $A(Y)>A(X)$), the two contributions can be resolved, and both the temperature of the gas and the line-of-sight turbulent motion directly estimated: 

\begin{equation} \label{eq:temp}
T = \frac{m_p}{2k} \frac{A(Y)A(X)}{A(Y) - A(X)} (b_{X^i}^2 - b_{Y^j}^2) = \frac{m_p}{2k} \frac{A(Y)A(X)}{A(Y) - A(X)} [(b_{X^i}^{th})^2 - (b_{Y^j}^{th})^2]. 
\end{equation}

\begin{equation} \label{eq:vel}
(\Delta \textrm{v})^2 = b_{X^i}^2 - [b_{X^i}^{th}(T)]^2. 
\end{equation}

The directly measurable quantities $b_{X^i}$ and $b_{Y^j}$, however, differ generally by amounts that are at least one order of magnitude smaller than the resolution of the current (and most of the currently foreseen) X-ray spectrometers (see Table \ref{tab:future_spec}), even for $\Delta \textrm{v} = 0$ (i.e. $b_{X^i} - b_{Y^j} = b_{X^i}^{th} - b_{Y^j}^{th}$) and for the largest possible difference in atomic weight. For example, for two absorption lines from carbon ($A(O)=12)$) and iron ($A(Fe)=52$) imprinted by gas at T$=10^6$ K,  $b_{X^i}^{th} - b_{Y^j}^{th} \simeq 14$ km s$^{-1}$. 

\noindent
Practically, thus, until X-ray spectrometers with spectral resolution better than a few tens to hundred km s$^{-1}$ will be available (e.g. the XGS of {\em Lynx}, see Table \ref{tab:future_spec}), one has to rely on ionization-balance estimate and modelling (\S \ref{sec:whim_ionbal}), to indirectly derive the temperature of the gas. 

\subsubsection{WHIM Chemical Conditions} \label{sec:whim_metal}
Finally, knowing the physical conditions of the gas (i.e. the ionization balance distribution $\xi(f)$),  one can estimate the relative abundances of the metals by their column density ratios. From Eq. \ref{eq:ioncol}: 
\begin{equation}\label{eq:rel_metal}
\frac{A(Y)}{A(X)} = \frac{N_{Y^j}}{N_{X^i}} \frac{f_{X^i}}{f_{Y^j}}.    
\end{equation}

These can then be compared to relative metallicity yields observed in different astrophysical environments and at different redshifts, to gain insights on the sources of metal pollution of the WHIM and the evolution of WHIM/CGM metallicity with cosmic time, or with the relative metallicities of the nearest galaxy to obtain hints on the galaxy-CGM/IGM feedback mechanisms. 

\subsection{Feasibility of LSS gas absorption Observations} \label{sec:feas}
Detecting X-ray absorption lines from gas in the LSSs, requires X-ray spectrometers with high resolution and throughput. 

Baryon densities in the range $n_b = 10^{-6} - 10^{-4}$ cm$^{-3}$ (over-densities $\delta \simeq 5-500$) and typical line of sight clouds (ranging from a fraction of galaxy's CGM - depending on the line-of-sight-to-galaxy impact parameter - to that of an entire WHIM filament oriented along line of sight direction) with sizes between $0.1-1$ Mpc, imply expected baryon column density of the order of N$_b \simeq 3\times(10^{17}-10^{19})$ cm$^{-2}$. For example, factoring a wide $Z = 0.01-1 Z_{\odot}$ range of metallicities, this yields, carbon, oxygen, neon and iron column densities N$_C \simeq (10^{12}-10^{16})$ cm$^{-2}$, N$_O \simeq 2\times(10^{12}-10^{16})$ cm$^{-2}$, N$_{Ne} \simeq 2\times(10^{11}-10^{15})$ cm$^{-2}$and N$_{Fe} \simeq (10^{11}-10^{15})$ cm$^{-2}$. The largest of these columns densities are expected from the densest WHIM environments, i.e. the nodes of the cosmic web, like outskirts of galaxy clusters or galaxy CGMs. 

Eq. \ref{eq:ew_lin} allows us to derive loose  upper limits (because of line saturation) to the rest-frame absorption line $EW$s expected from such ion column densities. He- and Ne-like ions are very stable both in purely photo-ionized or hybridly ionized gas, and their fractional abundance reaches values close to unity for wide ranges of ionization parameter or temperature (e.g. Figure \ref{f1_xtech}). Assuming unity, then, for the ion fractional abundance, yields to maximum observed $EW$s of C, O and Ne $He-\alpha$ and Fe $Ne-L\alpha$ transitions: $EW($\cv{$_{He-\alpha}$})$<30$ m\AA\ (0.2 eV), $EW($\ovii{$_{He-\alpha}$})$<60$ m\AA\ (1.6 eV), $EW($\neix{$_{He-\alpha}$})$<10$ m\AA\ (0.7 eV) and $EW($\fexvii{$_{Ne-L\alpha}$})$<25$ m\AA\ (1.4 eV). Lower ion fractional abundance and/or metallicity, as well as line-saturation, further reduce these upper limits.  

For a given absorption line with intrinsic width $b$ (in eV), the minimum $EW$ detectable in a spectrum with full-width-half-maximum resolution FWHM = $\Delta E$ (in eV) and signal to noise per resolution element $SNRE$ in the continuum, down to a single-line statistical significance N$_{\sigma}$ and neglecting instrument systematics, is given by the following formula: 

\begin{equation} \label{eq:min_ew}
    EW \ge 
    \begin{cases} 
    N_{\sigma} \frac{\sqrt{2.35 b  \Delta E}}{SNRE} \textrm{ eV}, & \textrm{if} (2.35 b) > \Delta E  \textrm{ (resolved lines)} \\ 
    &  \\
    N_{\sigma} \frac{\Delta E}{SNRE} \textrm{ eV}, & \textrm{if} (2.35 b) \le \Delta E  \textrm{ (unresolved lines)}\\
    \end{cases}
\end{equation}

Thus, a $\ge5\sigma$ detection of, e.g., an unresolved 10 m\AA\ (0.2 eV at 0.5 keV) WHIM absorption line at 0.5 keV (close to the \ovii{He-$\alpha$} transition) in an X-ray spectrum with FWHM resolution $\Delta E$, requires $SNRE \ge 25 \Delta E$ or,  considering only Poisson noise, $\ge 625 (\Delta E)^2$ counts per resolution elements. 

The number of counts per resolution element depends on the brightness $F$ of the background source, the throughput $A_{eff}$ of the spectrometer and the exposure $T$ of the observation. Thus, under the Poisson-noise-only assumption (i.e. neglecting systematics and other possible sources of non-Poissonian errors), we can write: 

\begin{equation} \label{eq:snre}
SNRE = \sqrt{10^{-3}\times F_E \Delta E A_{eff} T_{\rm exp}}, 
\end{equation} 

or 

\begin{equation} \label{eq:expo}
\frac{T_{\rm exp}}{10^3} = \frac{SNRE^2}{F_E \Delta E A_{eff}} \textrm{ ks}, 
\end{equation} 

where $F_E$ is the  specific photon flux at the energy $E$ in ph s$^{-1}$ cm$^{-2}$ keV$^{-1}$, $A_{eff}$ is the spectrometer effective area in cm$^2$, $\Delta E$ is the spectrometer resolution in eV and $T_{\rm exp}$ is the exposure time of the observation in s.  

Let us consider a bright local Seyfert, with a 0.5-2 keV flux $F = 2\times 10^{-11}$ erg cm$^{-2}$ s$^{-1}$ (1 mCrab). This, assuming for the 0.5-2 keV Seyfert spectrum a power-law with photon-index $\Gamma=2$ absorbed by a Galactic ISM column of $3\times 10^{20}$ cm$^{-2}$, translates into a 0.5 keV specific photon flux of $F_E \sim 0.032$ ph s$^{-1}$ cm$^{-2}$ keV$^{-1}$. 
Thus, from equations \ref{eq:min_ew}-\ref{eq:expo}, the minimum exposure time required to detect an unresolved 0.5 keV absorption line with $EW = 10$ m\AA\ (0.2 eV), against a 0.5-2 keV 1-mCrab background target and down to a single-line statistical significance of $\ge 5\sigma$, is 

\begin{equation} \label{eq:min_exp_10mA} 
T_{\rm exp} \textrm{\gax{}} 2\times 10^4 \left(\frac{N_{\sigma}}{5}\right)^2 \left(\frac{EW(eV)}{0.2}\right)^{-2} 
\left(\frac{F_E(ph\ s^{-1} cm^{-2} keV^{-1})}{0.032}\right)^{-1} \left(\frac{\Delta E(eV)}{A_{eff}(cm^2)}\right) 
%(625/0.032) [\Delta E(\textrm{eV})/A_{eff}(\textrm{cm}^2)] 
\textrm{ ks}. 
\end{equation} 

The XMM-{\em Newton} Reflection Grating Spectrometer (RGS) has 0.5 keV spectral resolution $\Delta E = 1.2$ eV and an effective area $A_{eff}(0.5 keV) \simeq 40$ cm$^2$, yielding to large single-target exposures $T_{\rm exp}$\gax{600} ks. For the {\em Chandra} Low Energy Transmission Grating (LETG) spectrometer, with a slightly better spectral resolution ($\Delta E = 1.0$ eV at 0.5 keV) but 4$\times$ lower effective area, $\sim$1 Ms exposures are required. 

This is why the search for WHIM absorption lines with the currently available high-resolution X-ray spectrometers, has proven to be extremely challenging and only produced few (and controversial (e.g. \citealt{Nicastro2005,Nicastro2005Nat,Kaastra2006,Rasmussen2007,Fang2002,Cagnoni2004,Fang2007,Fang2010,Ren2014,Nicastro2016c,Nicastro2018,Johnson2019}) detections over the 20-year period started with the advent of the first 2 high-resolution X-ray spectrometers. 

The situation will dramatically improve in the future (see \S \ref{sec:future}). The {\em Athena} X-Ray Integral Field Unit (X-IFU), with its $\Delta E = 2.5$ eV spectral resolution and 6,500 cm$^2$ effective area at 0.5 keV, will require only $T_{\rm exp}$\gax{7.5} ks observations to securely ($\ge 5\sigma$) detect the strongest ($\sim 10$ m\AA) WHIM lines against 1-mCrab targets. Much more common $EW=2-5$ m\AA\ (0.04-0.1 eV at 0.5 keV) WHIM lines will be detected against a 0.5-2 keV 1-mCrab target in $\sim 30-180$ ks and against 10 times fainter (and $\sim 100\times$ more numerous) targets, in $\sim 300-1800$ ks. This will open the route to the routine search and study of the diffuse gas in the LSS. 

\section{{Observations}} \label{sec:obs}

\subsection{Currently available instruments} \label{sec:gratings}
%(Chandra LETG/HERG and XMM RGS)

As seen in \S \ref{sec:feas}, intervening WHIM absorption lines are weak, requiring high spectral resolution in X-rays to detect them (\S \ref{sec:tech}). X-ray gratings became available with the launch of \chandra and \xmm in 1999, opening a new window in the studies of the diffuse warm-hot gas. There are two types of gratings available on \chandran: Low Energy Transmission Grating (LETG) and High Energy Transmission Gratings (HETG). The HETG has two grating arms, a Medium Energy Grating and High Energy Grating. The LETG can be used with the ACIS-S detector or with the HRC-S detector (where ACIS is the Advanced CCD Imaging Spectrometer, HRC is the High Resolution Camera, and -S is for the spectroscopy mode).  The \xmm observatory has two Reflection Grating Spectrometers (RGS1 and RGS2). In Table \ref{tab:future_spec} we provide the basic properties of these gratings. 

\subsection{Intervening X-ray absorption lines} \label{sec:whim}

 With the newly deployed X-ray gratings on \chandra and \xmmn, astronomers could finally look for the missing baryons in the low-redshift WHIM. However, attempts to detect $z>0$ absorption lines were fraught with difficulties. 
 While doing a blind search for intervening absorption lines, we have to pay a statistical penalty for chance fluctuations that may appear as lines. For example, HRC-LETG has 1,000 resolution elements in the $10$--$60$\AA\ region. Therefore, there may be three ($(1-0.997)\times 1,000$) spurious ``lines'' in the spectrum with statistical significance of $>3\sigma$ (see e.g. \citealp{Das2021b}). To counter this problem, often some redshift markers from ancillary observations (e.g. \ovi absorbers in the UV) are used so that the location of the X-ray line is known {\it a priory} \citep{Mathur2003}. 
 
 In order to detect line-of-sight absorption lines, we need a bright background source. In the UV regime, bright quasars are used for this purpose. However, most quasars (particularly radio-quiet quasars), have intrinsic absorption lines. The high spectral resolution in the UV allows for easy separation of intrinsic absorption from the intervening absorption lines. This, however, becomes extremely difficult with current generation of X-ray grating spectroscopy. Therefore it is necessary to use blazars as background X-ray sources, as they have featureless continua. 
 
 With these considerations, efforts to detect the intervening X-ray absorption lines have been toward blazar sightlines (discussed below). Nevertheless, these have been largely unsuccessful, barring some notable exceptions. Even in the cases where the lines were reported, it was not clear whether they arise in the WHIM, or in the circumgalactic medium (CGM) of external galaxies, or in the intra-group medium. Below we discuss the  efforts to detect the $z>0$ X-ray absorption lines as reported in literature. 
 
\subsubsection{Sightline to H\,1821+642}\label{sec:H1821}

\begin{figure}
\centering
\includegraphics[trim=20 400 10 10,clip,width=12.0cm]{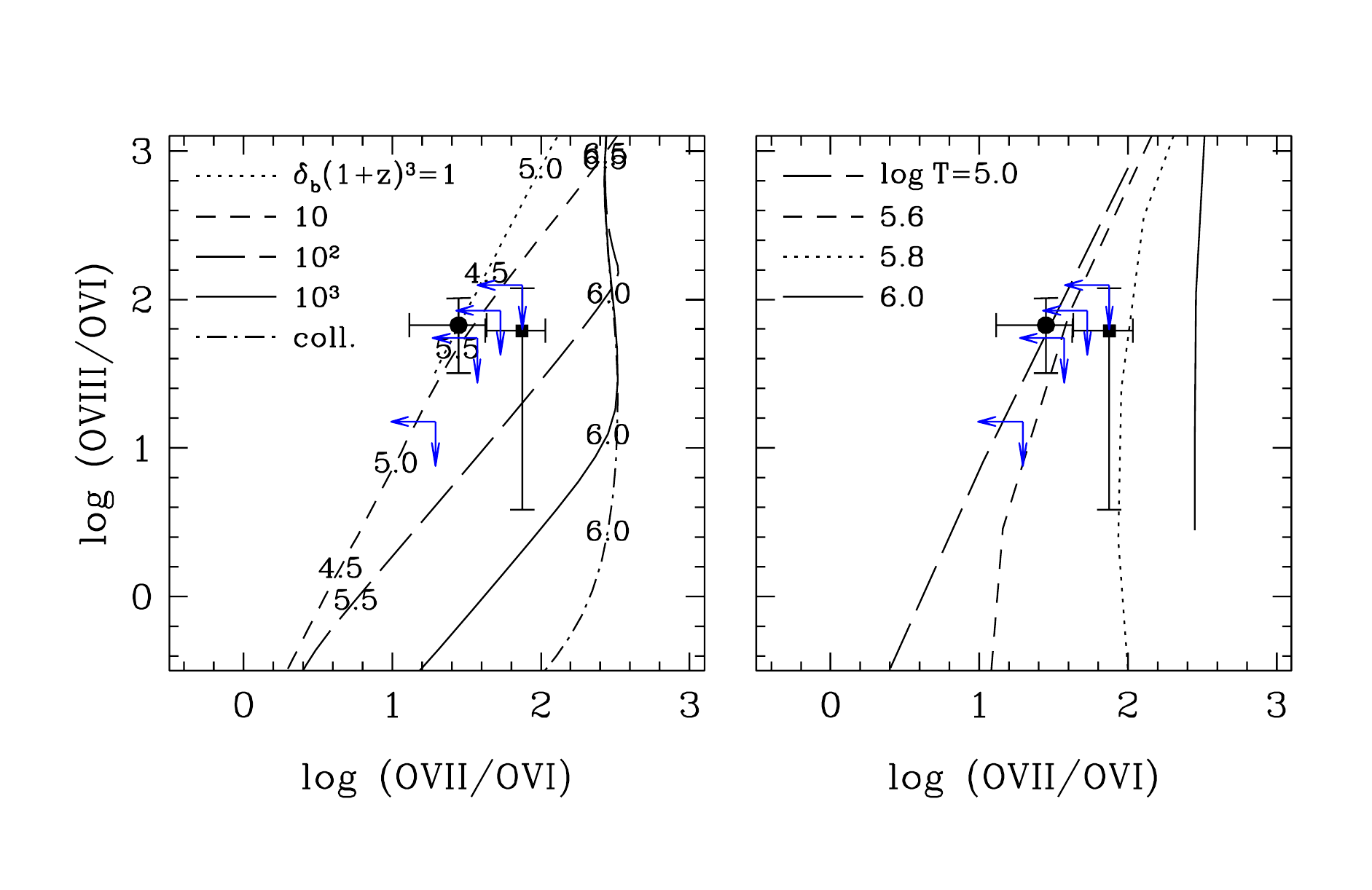}
\caption{Constraints on the physical state of the intervening absorbers with \ovin, \oviin, and \oviii: black points show the two tentative \ovin-associated X-ray detections, while blue upper limits refer to the 4 \ovin\ absorbers not detected in the X-rays. 
Left: Curves show the tracks in the 
$f$(\oviiin)$/f$(\ovin) vs.\ $f$( \oviin)$/f$(\ovin) plane. Dotted, short-dashed, long-dashed,
and solid lines are for gas overdensities $\delta_b(1+z)^3=1$, 10, $10^2$,
and $10^3$, respectively, while the dot-dashed line represents pure
collisional ionization.  Numbers
along these curves indicate $\log T$ in degrees Kelvin. 
  Right: Same, but with tracks
of constant temperature. Figure adapted from \cite{Mathur2003}. 
}
\label{fig:mathur}
\end{figure}

A major effort to detect the WHIM with \ovii absorption lines was made by \citet{Mathur2003} along the sightline to the blazar H\,1821+642, using intervening \ovi lines to select the absorber redshifts. Conclusive evidence for the WHIM detection was not reported by these authors, but their analysis provided ground work for future observations. They showed the power of using \oviin, \oviii lines in the X-ray band, together with the \ovi line in the UV.  In Figure \ref{fig:mathur}, tracks of  theoretical models are plotted in the $f$(\oviiin)$/f$(\ovin) vs.\ $f$(\oviin)$/f$(\rm \ovin) plane, where, e.g. $f$(\oviin) is the ionization fraction of \oviin. On the left, the tracks are for different gas overdensities, while on the right they are for different temperatures. We see that $f$(\oviin)$/f$(\ovin) is primarily a diagnostic of gas
temperature, while $f$(\oviiin)$/f$(\ovin) constrains the gas density
for a given value of $f$(\oviin)$/f$(\ovin).  This behavior reflects
the competing roles of photoionization and collisional ionization,
with the latter being more important for higher temperatures,
higher densities, and lower ionization states (see also \S \ref{sec:abs_obs} and Fig. \ref{f2_xtech}).

%-----------------------------Figure Start------------------------------
%-----------------------------Figure End--------------------------------

\citet{Kovacs2019} stacked the \chandra spectrum around the wavelengths of the \ovii lines at the redshifts of 17 intervening \lya\ lines and detected the \ovii line with $3.3\sigma$ significance. They found the \ovii column density of $N_{\rm OVII}=1.4\pm0.4 \times 10^{15}$ cm$^{-2}$. Assuming that all the 17 systems contribute equally, they determined the gas density to be $n_H=1-2 \times 10^{-6}$ cm$^{-3}$, corresponding to the density in the WHIM. However, theoretical simulations show that while \ovi correlates with \ovii and \oviii in the WHIM, the same is not true for \hi\ \citep{Wijers2019}. Therefore it is unlikely that stacking on \hi \lya\ yields strong \ovii absorption. \citet{Kovacs2019} selected only those foreground \lya\ absorbers that were associated with massive galaxies with $M_{\star}$\gax$10^{10}$\msun, corresponding to $M_{\rm halo}$\gax$3\times 10^{11}$\msun. It is therefore possible that the \ovii absorption is dominated by the CGM of five of their galaxies with impact parameters within twice the virial radii.

\subsubsection{Sightline to Mrk\,421}\label{sec:mrk421}

\begin{figure}
\centering
\includegraphics[trim= 350 150 10 150,clip,width=25.0cm]{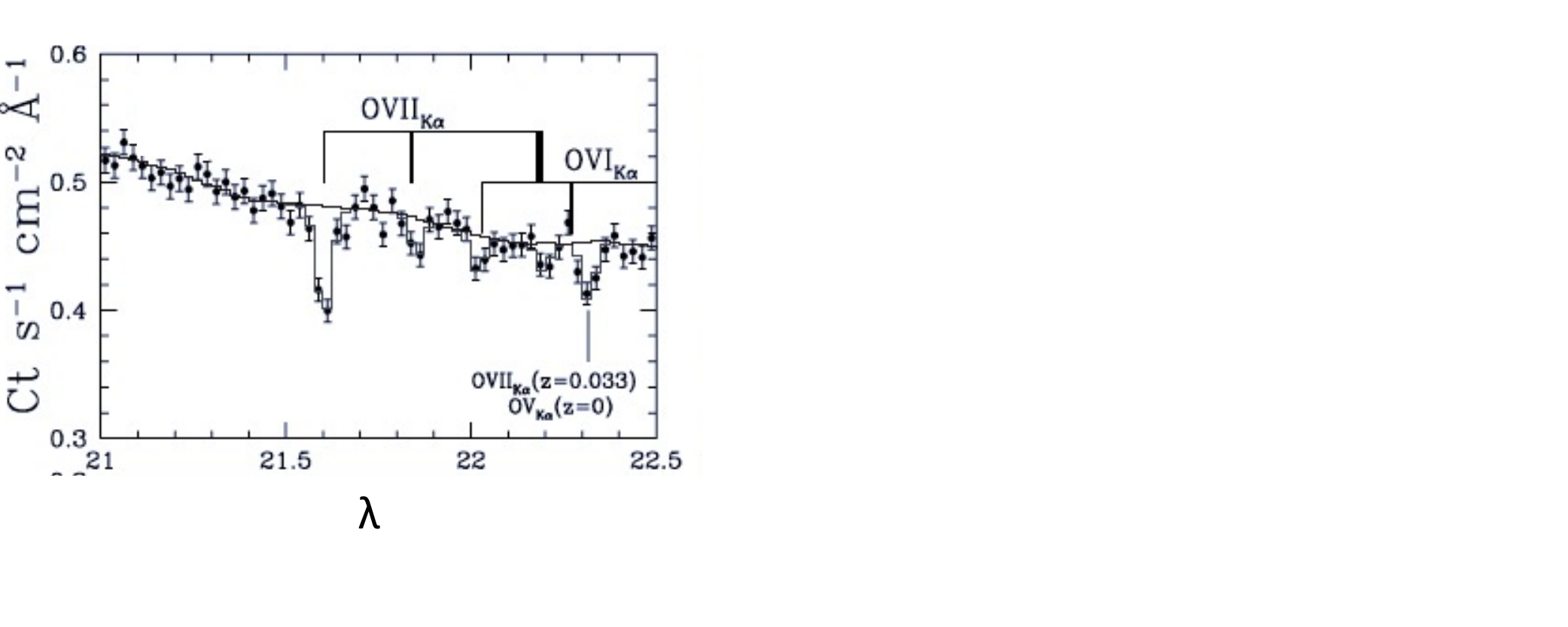}
\vspace*{-1cm}
\caption{The \chandra spectrum towards Mrk\,421 (from \citealp{Nicastro2005}). Note the \ovii K$\alpha$ lines at $z=0$ and at intervening redshifts of $z=0.011$ and $z=0.026$, marked with thin, thick, and very thick bars.  
 }
\label{mrk421}
\end{figure}

\citet{Nicastro2005} reported detection of WHIM lines in the sightline to Mrk\,421. In order to obtain a high SNRE spectrum, the authors monitored the blazars in the sky, and triggered \chandra observations when the source was about ten times brighter than the average. The resulting spectrum (Figure \ref{mrk421}) led to the detection of two intervening systems at redshits $z=0.011\pm0.001$ with the \ovii \ka\ absorption line and at $z=0.026\pm0.001$ with \nvii \ka\ line.

Whether these systems are from the WHIM or from outskirts of galaxies is a matter of debate \citep{Williams2010}. 
Using the highest SNRE \chandra spectra, \cite{Das2021b} did not detect the \ovii absorption lines reported by \cite{Nicastro2005} (see \citealt{Rasmussen2007} for non-detections in \xmm data), but they did detect the \oviii \ka\ lines at the two redshifts, consistent with the upper limits in \citet{Nicastro2005}. Thus the detection and interpretation of the intervening absorption toward Mrk\,421 remains inconclusive. 

%-----------------------------Figure Start------------------------------
%-----------------------------Figure End--------------------------------

\subsubsection{Sightline to PKS\,2155-304}\label{sec:pks2155}

The first reported detection of the WHIM is along this sightline \citep{Fang2002}. There are both \ovi and broad \lya\ intervening absorbers detected, providing good anchors for searching for X-ray lines. The most recent analysis of PKS\,2155-304 spectra using \chandra and \xmm is presented in \citet{Nevalainen2019}, and the results are confusing. The deep  \xmm spectra did not result in the detetion of any  absorption from the intervening systems. \chandra LETG/HRC spectra did not show any detected line either. However, the \chandra LETG/ACIS-S data yielded a line-like feature at $3.7\sigma$ which could be an \oviii absorption line at the intervening redshift of $z=0.054$. The \chandra results are consistent with those reported previously \citep{Fang2002,Fang2007,Williams2007}. While the detected line could be from the WHIM, the authors could not rule out that it could also be a transient feature associated with the blazar. Moreover, the detection itself is not robust given the inconsistency among different detectors. 

\subsubsection{Sightline to 3C\,273}\label{sec:3c273}

Absorption signatures along the sightline to this blazar were explored by \citet{Fang2003} using \chandra data. They reported on the detection of the \ovii absorption line at $z=0$, but not from the intervening WHIM. Recently, \citet{Ahoranta2020} explored the same sightline with \xmm and \chandra and reported the detection of \ovii and \neix lines from the WHIM  at $z=0.09017$ where a \ovi system is detected in the UV. While they claim a $3.9\sigma$ combined significance of the detection, inspection of their spectra shows that the evidence is marginal. Moreover, as noted above in \S \ref{sec:whim}, there is a high probability of spurious lines with low statistical significance. Thus the presence of the WHIM detection in this sightline remains doubtful. 

\subsubsection{Sightlines to H\,2356-309 and Mrk\,501}\label{sec:H2356}

The sightline to blazar H\,2356-309 is interesting because it passes through the Sculptor Wall, a LSS of galaxies. The redshift of the LSS is known ($z\sim 0.03$), providing a marker for X-ray absorption lines. \chandra and \xmm observations of this sightline were presented by \citet{Fang2009} who reported a $3\sigma$ detection of the \ovii \ka\ line. A deeper \chandra data yielded a $3.4\sigma$ line detection, and the significance increases to $4\sigma$ including the \xmm data \citep{Fang2010}. A similar detection behind the Herculus wall ($z\sim 0.03$) in the sightline toward Mrk\,501 was reported by \citet{Ren2014}. However, both of these line detections from the putative WHIM (at the same redshift) were likely contaminated by \oii\ k$\beta$ lines from the interstellar medium of the Milky Way \citep{Nicastro2016b}. 

\subsubsection{Sightline to 1ES\,1553+113}\label{sec:1es}

\begin{figure}
\centering
\includegraphics[trim=130 10 0 10,clip,width=14.0cm]{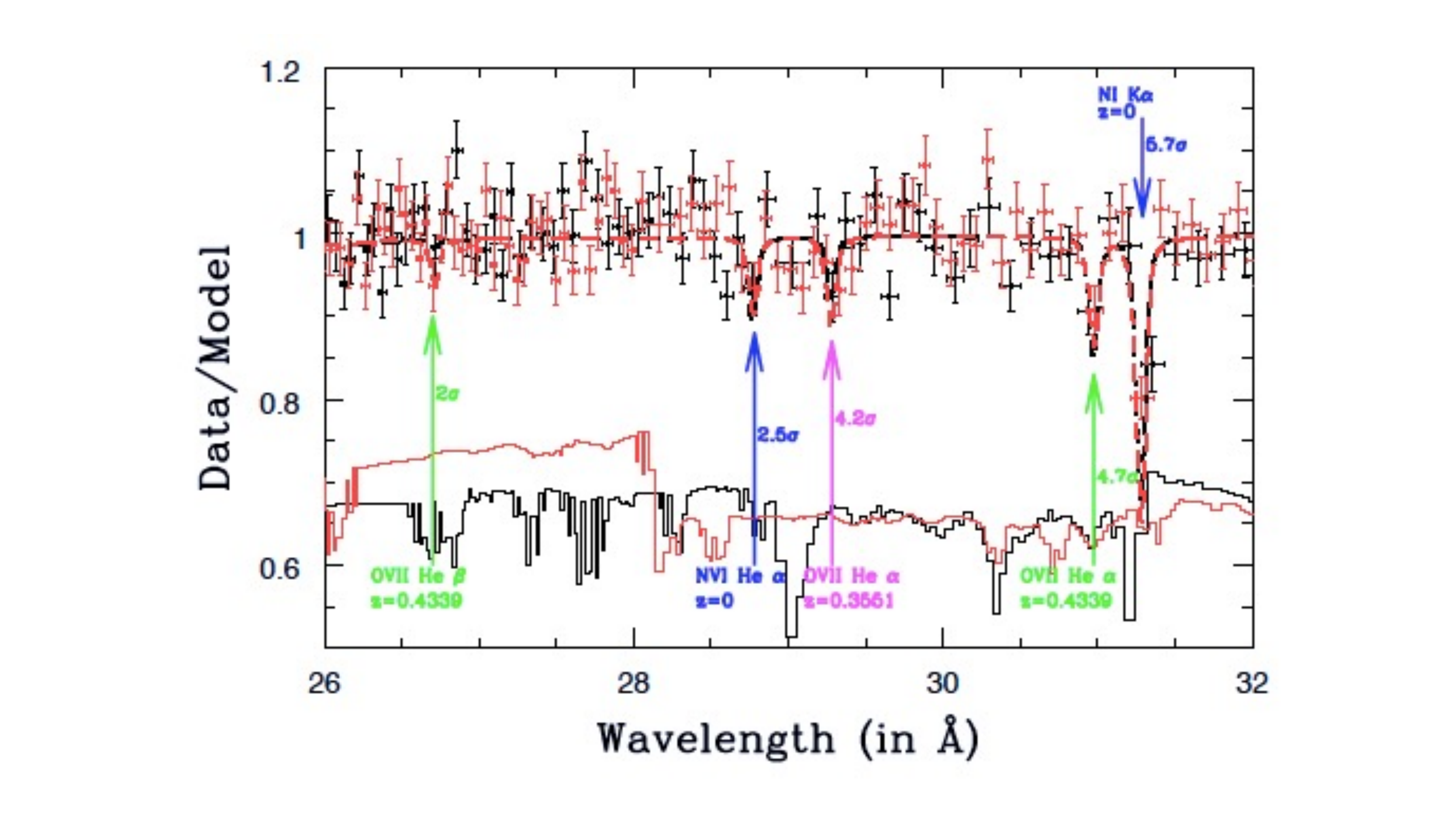}

\caption{{\it Left}: The \xmm spectrum towards \1es (from \citealp{Nicastro2018}) reporting the detections of the intervening absorption lines at $z1=0.4339$ and $z2=0.3551$. Shown are the normalized raw data (in bins with signal to noise per bin $\geq 30$) in RGS1 (black points and $1\sigma$ errorbars) and RGS2 (red points and $1\sigma$ errorbars). Thick dashed curves are RGS1 (black) and RGS2
(red) best-fitting model folded through the RGS response functions. Thin solid
curves at the bottom of the graph are RGS1 (black) and RGS2 (red) effective areas
(in arbitrary units), showing instrumental features due to cool-pixels in the
detectors. Of the 5 absorption lines shown, only the weak \ovii \kb\ at
$z=0.4339$, and only in the RGS1, can be affected by the presence of an instrumental
feature.
 }
\label{fig:1es}
\end{figure}

From the experience of observing several  sightlines over the years, it was clear that  deep observations of a blazar sightline are necessary to detect the WHIM with current generation of instruments. The sightline to \1es was observed with \xmm for almost $1.8$Ms by \citet{Nicastro2018}. The resulting spectrum reached SNRE$=33$ in the $23.5$--$30.2$\AA\ spectral region (shown in Figure \ref{fig:1es}). The high SNRE allowed detections of weak absorption lines with $EW$\gae$3.5$ m\AA. The authors reported the detections of two intervening absorption systems at $z1=0.4339$ and $z2=0.3551$. The $z1=0.4339$ system was detected with the X-ray lines of \ovii \ka\ and \kb.  The $z2=0.3551$ system was detected with the \ovii \ka\ X-ray line only. The existence  of these systems was further corroborated by the presence of galaxy overdensities around the sightline at the absorber redshits. There is a broad \hi \lya\ absorber at the redshift of the $z1$ system. There are two \hi \lya\ absorbers close to the $z2$ system, though they are separated by $750$\kms.

However, \citet{Johnson2019} question whether the intervening absorption systems are truly from the WHIM. These authors  present a deep redshift survey in the field of \1es. They found that the blazar, whose spectroscopic redshift is still unknown, is likely a member of a group of galaxies, and that the $z1$ system arises in the intra-group medium. If the $z1$ system is associated with the LSS associated with the blazar, it should not be included in the statistics of the blind search. They also show that the $z2$ system arises in an isolated environment which they show to be unexpected for the WHIM. Therefore, \citet{Johnson2019} argue that the X-ray intervening systems do not trace the WHIM, and do not account for the missing low-redshift baryons (see \S \ref{sec:missb}).  

%-----------------------------Figure Start------------------------------
%-----------------------------Figure End--------------------------------

\subsection{WHIM and the CGM}

In the discussion above, we have been discussing the CGM alongside the WHIM. There are several reasons why it is difficult to distinguish between the WHIM and the CGM. Both are observed as intervening absorption systems, and unless we know of the presence of nearby galaxies, we cannot tell whether or not they arise within the CGM. Theoretical simulations do not necessarily distinguish between the two either. Both the WHIM and the CGM have $T\approx 10^6$K, and the gas is diffuse. The $T$--$n$ phase space in simulations, however, shows a wide range of densities at  $T\approx 10^6$K (Figure \ref{BertoneMetallicity}) with denser gas belonging to the CGM while the truly diffuse gas (with density down to almost cosmological mean density) corresponding to the WHIM. However, with absorption lines we measure the column density, not the density, leading to the ambiguity in the inferences. New simulations are providing insights on linking density to column density. Using the EAGLE cosmological simulations, \citet{Wijers2019} studies the column density distribution functions (CDDF) of intervening \ovii and \oviii absorption systems (see \S \ref{sec:IGM}). They found a break in the CDDF around the column density of $\rm \log N$(\oviin)$=10^{16} cm^{-2}$ dominated by systems with overdensities of $\sim 100$. Therefore, absorption systems with column densities lower than the break likely arise in the diffuse portion of the WHIM, while those with higher column densities are from the CGM of collapsed halos (see Figure \ref{fig:ccdf2}). The absorption systems detected so far (\S \ref{sec:whim}) are therefore likely from the CGM of external galaxies (or from the intra-group medium (\S \ref{sec:1es})). \\

Another distinguishing characteristic between the dense (CGM) and the diffuse WHIM is the metallicity. The CGM is likely to be more metal-rich than the WHIM. Observationally it is difficult to measure the metallicity directly from X-ray observations alone. However, a combination of UV and X-ray spectroscopy would allow metallicity measurements and further distinguish the CGM and the WHIM. The CGM is discussed in detail in the Chapter ``Probing the circumgalactic medium with X-ray absorption lines'' of Section XI of this Handbook. 
 
\subsection{WHIM and the missing baryons}\label{sec:missb}

\begin{figure}
\centering
\includegraphics[trim=400 0 350 0,clip,height=12.0cm]{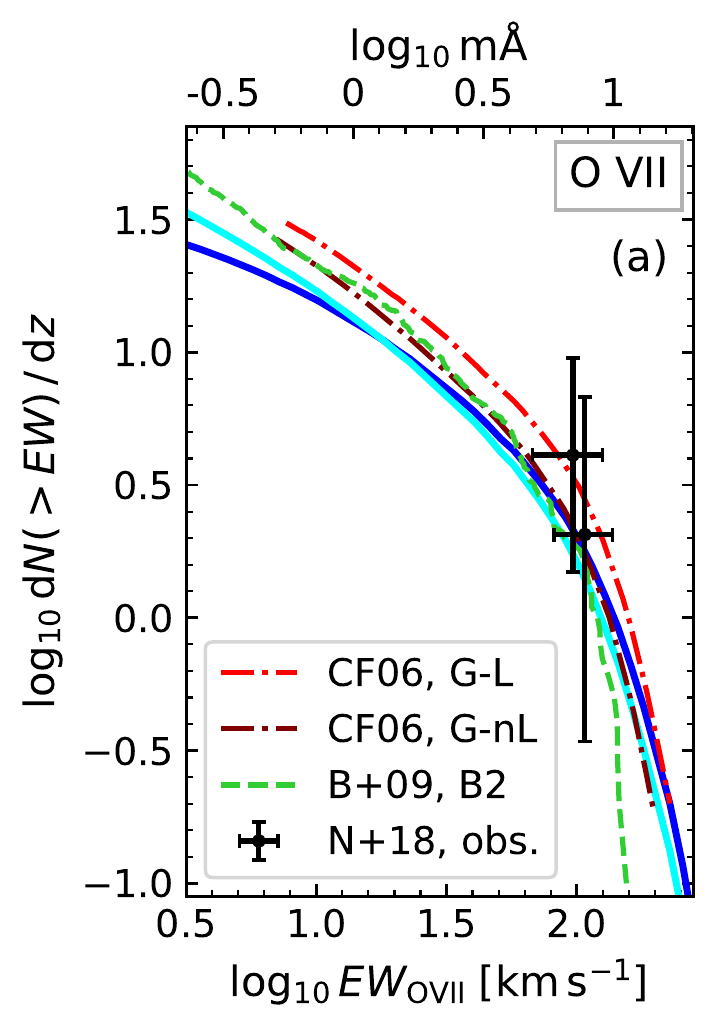}
\caption{The distribution function of \ovii EWs (from \citealp{Wijers2019}). The blue and cyan lines are from their EAGLE simulations; the dot-dashed lines are from the simulations of \citet{CenFang2006}; and the green dashed line is from \citet{Branchini2009}. The two points with error-bars are from \citet{Nicastro2018}. 
 }
\label{fig:ccdf2}
\end{figure}

The intervening X-ray absorption lines from the WHIM allow us to determine the baryon content of the WHIM. Indeed, from the measured column density of the \ovii ion one can obtain the baryon density of the low-redshift Universe as follows. Converting the \ovii column density to the equivalent total column density depends on the temperature, oxygen abundance and metallicity of the WHIM. They are not directly measured with X-ray spectroscopy; therefore assumption are made, guided by theory (see \S \ref{sec:theory}). The next step is to convert the observed number density of the absorbers (number per unit redshift/pathlength) along a few sightlines to the cosmic baryon density. While this is an established method in the UV, we again need to resort to theoretical simulations in the X-ray band. This is because there are only a few sightlines with detections of intervening absorption, resulting in large errors on the number density of the absorbers. Secondly, the observations do not span the wide range of column densities to determine their distribution function empirically, so once again we need to resort to theoretical models. Additionally, the detected systems lie on the high end of the distribution function predicted by theoretical models (see Figure \ref{fig:ccdf2}). As noted in \S \ref{sec:1es} above, the lines may not be from the true WHIM, but from the CGM of nearby galaxies or from the intra-group medium, in which case they do not contribute to the column  density distribution function of the WHIM. As such, there are many caveats in determining the observed baryon fraction in the WHIM from the X-ray detected gas. 

\begin{figure}
\centering
\includegraphics[trim=0 0 0 0,clip,width=12cm]{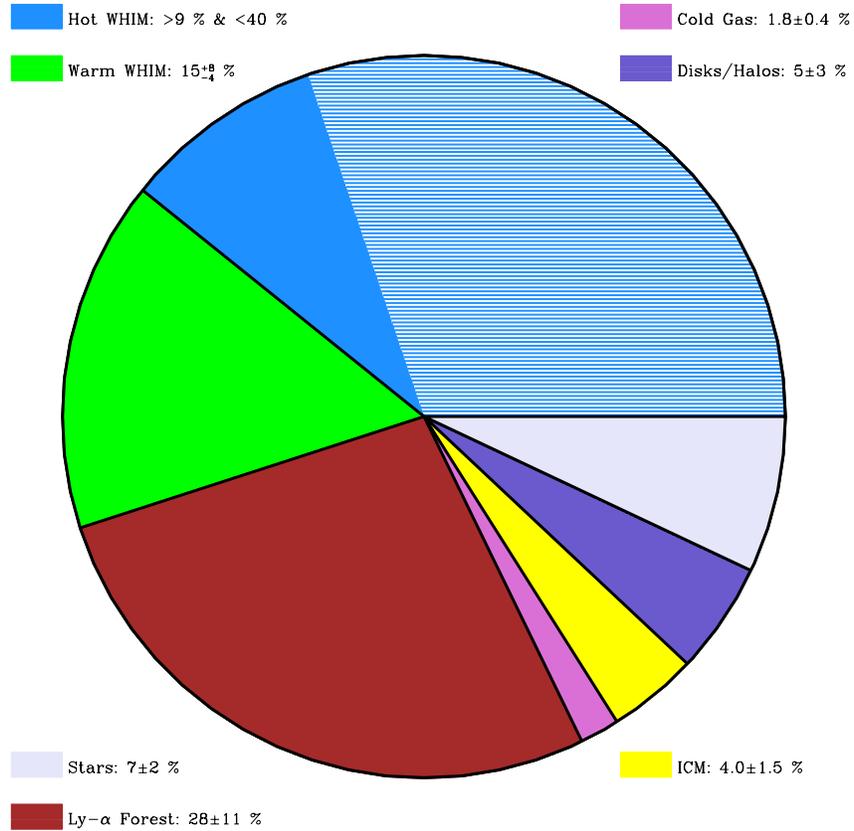}
\vspace*{-2.0cm}
\caption{
Pie diagram of the baryonic components of the local Universe (from \citealp{Nicastro2018}). 
 }
\label{fig:baryoncensus}
\end{figure}

Nevertheless, assuming that the two detected intervening absorption systems along the \1es sightlines are true detections, and that they are from the WHIM, and assuming that the extrapolation of the column density distribution function to lower columns is as predicted by theory, one can measure the cosmic baryon census (see Figure \ref{fig:baryoncensus}).  According to \citet{Nicastro2018}, the warm-hot gas traced by X-ray absorption lines can account for the missing baryons in the low-redshift Universe.

%-----------------------------Figure Start------------------------------
%-----------------------------Figure End--------------------------------
 
%-----------------------------Figure Start------------------------------
%-----------------------------Figure End--------------------------------

\section{\textit{Future}} \label{sec:future}
%Athena, Arcus, Lynx + atomic database
The future of LSS-absorption studies is bright. Several new X-ray missions (or mission concepts) have been studied and designed over the past 2 decades to overcome the limitations of current high-resolution X-ray spectrometers (namely, small effective area and relatively low spectral resolution: see \S \ref{sec:gratings} and \ref{sec:feas}) and some of these have been selected by space-agencies and are currently planned (or likely) to fly over the next two decades, such as \athena{} \citep{barcons12_athena}, \xrism{} \citep{Tashiro18_xrism}, \arcus{} \citep{Smith20_arcus}. 
%Table \ref{tab:newmissions} summarizes the property of these proposed missions and shows a figure of %merit of their high-resolution spectrometers, relatively to LSS absorption studies. 

Modern, high-efficiency high-resolution X-ray spectrometers fall into two broad categories: dispersive and non-dispersive. In the following we schematically review the main properties, as well as advantages and disadvantages of these two types of spectrometers, for WHIM studies. 

\subsection{\textit{Dispersive Spectrometers}} \label{sec:disp_spec}
Dispersive X-ray spectrometers use X-ray gratings (see Section I of this book, Chapter {\em Gratings for X-ray Astronomy}) to diffract (in one dimension) photons onto an array of detectors with dispersion angles $\beta$ that depend linearly on the wavelength $\lambda$ of the incoming photons: 
\begin{equation} \label{eq:disp_spec} 
sin\beta = m \frac{\lambda}{p},  
\end{equation}

where the integer $m$ (0, 1, 2,...) is the order number and $p$ is the spatial period of the grating lines. 
Non-diffracted photons ($\beta=0$, or 0$^{\rm th}$-order) form an image of the source on the detector, while higher-order photons are dispersed on the detector to a distance from the centroid of the 0$^{\rm th}$-order image proportional to the order number and the energy of the photons. Thus, the energy resolution of gratings depends on the precision with which the photons are localized on the detectors (and so, in turn, on the spatial resolution of the mirror along the dispersion direction). Different orders, superimposed on the same dispersion axis, are separated by exploiting the (much lower) energy resolution of the detectors. 

Because of the technique used to measure the energy of the incoming photons (eq. \ref{eq:disp_spec}), the ``natural" unit for grating spectrometers is the photon wavelength, in \AA. 

Currently operating grating spectrometers onboard \chandra{} \citep{Weisskopf99_chandra} are transmission gratings (i.e. interposed between the mirror and the focal plane, along the focal axis) while those onboard \xmm{} \citep{Jansen01_xmm} are reflection gratings (i.e. placed in the converging beam at the exit from the X-ray telescopes and reflecting photons to detectors offset from the telescope focal plane). These instruments have been designed several decades ago and are characterized by low diffraction efficiency (a few percent at 1 keV) and 1$^{\rm st}$-order resolving powers of a few hundreds to a thousand over the covered spectral ranges (see Table \ref{tab:future_spec}). 

New grating spectrometers with much improved efficiency (up to 60-70\% at 1 keV: e.g. \citealp{Heilmann21_catgrat}) and constant resolving power over the covered energy band (obtained by exploiting different diffraction orders with enhanced intensity in different portions of the spectrum) of several thousands, have been developed over the past decade, and have reached high technical readiness levels (e.g. \citealp{Heilmann21_catgrat, Smith20_arcus}). 
These gratings have been proposed for several mission concepts (Table \ref{tab:future_spec}), e.g. WhimEx \citep{Cash11_whimex}, \arcus{} \citep{Smith20_arcus}, \lynx{} \citep{Gaskin_lynx19}, \hirex{} \citep{Nicastro21_hirex}, and one of them (\arcus{}) was  selected by NASA in 2016 for a phase-A concept study. 

Dispersive spectrometers can reach today resolving powers $R$ of several thousands and up to R$\simeq 10,000$ and over (e.g. \citealp{Gunter_Heilmann19,Nicastro21_hirex}), thus approaching the resolution needed to directly measure the temperature of the absorbing gas by comparing line-widths imprinted by ions with different atomic weight (see \S \ref{sec:t_and_kin}). 
However, these types of high-resolution X-ray spectrometers allow only efficient spectroscopy of one target at a time (located in the proximity of the central region of the detector), and are optimally suited for point-like (i.e. non-spatially resolved) sources. 

\subsection{\textit{Non-Dispersive Spetrometers}} \label{sec:non-disp_spec}
To enable simultaneous spectroscopy of all (extended and point-like) sources present in the field of view (Integral-Field-Unit spectroscopy), arrays of non-dispersive high-resolution spectrometers (each representing a pixel of the detector) are needed in the focal plane of the mirror. 

X-ray micro-calorimeters have been developed over the past two decades, and consist of essentially three elements: an absorber, a resistive thermometer and a heat-sink (or cooler). 
%Micro-calorimeters work by measuring the change in resistance caused by the $\sim$milli-kelvin raise in temperature (compared to the cooling-temperature guaranteed by the heat-sink) generated in the absorber by the ejection of a photo-electron due to absorption of an X-ray photon by the atoms of the absorber's material. The cooler is needed because the energy resolution of these devices scales with temperature, so cryogenic temperatures are needed for high-energy resolution.

The most developed X-ray micro-calorimeters are called Transition-Edge-Sensors (TESs: see Section II of this book, Chapter {\em TES detectors (excluding fabrication)} for details).  
%and exploit the superconducting property of metals at their transition temperature $T_c$ characteristic of the transition from superconductor to normal metal. An important parameter of TESs, is the so called $\alpha$ parameter, which measures the ability of the material (at given temperature $T$ and resistance $R$), to respond to temperature variations with changes in resistance, and is defined as: 
%\begin{equation} \label{eq:tes_alpha} 
%\alpha = \frac{T}{R} \frac{dR}{dT},  
%\end{equation}
The energy resolution of a TES is inversely proportional to the square root of a parameter $\alpha$ that measures the ability of a material to respond to temperature variations with changes in resistance, and can be written: 
\begin{equation} \label{eq:tes_deltae} 
\Delta E \simeq 2.35 \sqrt{4kT^2\frac{C}{\alpha}},  
\end{equation}

where $C$ is the TES's heat capacity. Thus, the lower the operating temperature of the absorber and its heat-capacity, and the higher the parameter $\alpha$, the better the energy resolution. There are intrinsic limitations, however. TESs can only operate near their transition temperature $T_c$. 
%Far from $T_c$ the superconductor goes normal, and $\alpha \sim dR/dT \to 0$. 
At a given operating temperature, this implies a maximum $\Delta T$ in the TES, which in turn implies a maximum energy of the incoming photons: 
\begin{equation} \label{eq:tes_emax} 
E_{MAX} \simeq C \Delta T_{MAX} \simeq C \frac{T}{\alpha},  
\end{equation}

And finally, replacing Eq. \ref{eq:tes_emax} in \ref{eq:tes_deltae}, we get: 
\begin{equation} \label{eq:tes_deltae1} 
\Delta E \simeq 2.35 \sqrt{4kTE_{MAX}},  
\end{equation}

Thus the resolution of a TES operating at 100 mK, cannot be better than $\simeq 1$ eV at 6 keV and $\simeq 0.4$ eV at 1 keV. 

Several future (or planned) X-ray missions will have an array of TESs in their focal plane, e.g. \xrism{} (JAXA/NASA mission, planed launch-date year 2022; \citealp{Tashiro18_xrism}) \athena{} (ESA large-mission, launch-date currently foreseen in year 2031; \citealp{barcons12_athena}), \lynx{} (NASA mission concept, currently under decadal-review; \citealp{Gaskin_lynx19}), \hubs{} (China National Space Administration - CNSA - mission concept, currently under study; \citealp{Cui20_hubs}). 

\subsection{\textit{Detectability and Study of LSS Absorbers with Future Missions}} \label{sec:future_lssabs}
As shown in Eq. \ref{eq:min_ew} (\S \ref{sec:tech}), detecting absorption lines from gas in the LSS down to a given line EW  
%and inferring information on its physical, kinematical and chemical states, 
requires large throughput and high-spectral resolution in the 0.2-2 keV band. 
We therefore define the square root of the product between the resolving power and the effective area of the detector at 0.5 keV, as a figure-of-merit for LSS absorption detectability (FoM$_{det}$). 
Table \ref{tab:future_spec} summarizes the properties of the main, planned or under concept studies, future X-ray spectrometers and show their FoM$_{det}$, normalized to that of the current \chandra{} HRC-LETG spectrometer. 

\begin{table}
%\scriptsize
\begin{center}
\caption{Current \& Future high-resolution X-ray spectrometers}
\label{tab:future_spec}
\end{center}
\begin{tabular}{|ccccccc|}
\hline
Mission   &  Spectrometer & Instrument & Effective Area & Resolving Power & Spectral Resolution & FoM$_{det}$ Relative \\
    &              &  & at 0.5 keV (in cm$^2$)  & at 0.5 keV & at 0.5 keV (in km s$^{-1}$) & to \chandra{LETG} \\
\hline
\chandra{} & HRCS-LETG & Grating & $^c$12 & 496 & 604 & {\bf 1} \\
\xmm{} & RGS1+RGS2 & Grating & $^a$40 & 382 & 785 & {\bf 1.6} \\
\hline
\xrism{} & Resolve & TES & 65 & 83 & 3612 & {\bf 0.95} \\
\athena{} & X-IFU & TES & 5900 & 200 & 1500 & {\bf 14.1} \\
\hubs{} & XQSC & TES & 450 & $>$500 & $<$600 & {\bf $>$6.1} \\
\hline 
\arcus{} & & Grating & 800 & 2500 & 120 & {\bf 18.3} \\
\lynx{} & XGS & Grating & $>4000$ & $>5000$ & $<$60 & $>${\bf 58} \\
\hline
\end{tabular}
\noindent
a. First order \\
\end{table}

For each of these missions, systematic effects (uncalibrated pixel-to-pixel quantum efficiency variations) in the focal plane detector are the ultimate limit to the minimum detectable line equivalent width. For an assumed 3\% systematic uncertainty, the minimum detectable equivalent width is 3\% of the instrument resolution. Table \ref{tab:future_detectabillity} lists these minimum EWs (EW$_{min}$: Col. 4) for lines at 0.5 keV, together with the corresponding median column of \ovii{ K$\alpha$} (N$_{OVII}^{min}$: Col. 5), as derived from EAGLE hydro-dynamical simulations (Figure 7 in \citealp{Wijers20}). 
We also use Eq. \ref{eq:min_exp_10mA} to derive the minimum exposure time (in Ms) needed to detect such LSS absorption lines at 0.5 keV down to the their systematic limit EW$_{min}$ and against background-targets with 0.5-2 keV flux $f_E=1$ mCrab ($2\times 10^{-11}$ erg s$^{-1}$ cm$^{-2}$; Table \ref{tab:future_detectabillity}, Col. 6). 
There are about 15 of such targets (AGNs, most of which blazars) in the extragalactic X-ray sky, covering a total line-of-sight redshift interval of $\Delta z = 7.9$. 
In Table \ref{tab:future_detectabillity} (Cols 8 and 9), we therefore list also the total exposure (in yrs) needed to cover this redshift pathlength (Col. 7), together with the expected number of LSS absorbers per unit redshift down to the detectability limit N$_{OVII}^{min}$ (Figure \ref{WijersCDDF}) and the total number of detectable LSS absorbers in the full $\Delta z=7.9$ pathlength. 
Finally, we define a second Figure of Merit for LSS absorber studies (FoM$_{std}$), given by the number of detectable 0.5 keV absorbers per year (ratio of Col. 10 to 9).  

\begin{table}
%\scriptsize
\begin{center}
\caption{Maximum Detectabillity of LSS absorbers with current and future Missions}
\label{tab:future_detectabillity}
\end{center}
\begin{tabular}{|cccccccccc|}
\hline
Mission   &  Spectrometer & Instrument & EW$_{min}$ & N$_{OVII}^{min}$ & T$_{min}$ & T$_{min} \times 15$ & Systems & Total Number & FoM$_{std}$ \\
    &              &  & at 0.5 keV (in eV) & (in $10^{15}$ cm$^{-2}$) & (in Ms) & (in yrs) & per unit z & of Systems & (in yr$^{-1}$) \\
\hline
\chandra{} & HRCS-LETG & Grating & 0.03 & 0.5 & 27 & 13 & 15.8 & 125 & {\bf 10} \\
\xmm{} & RGS1+RGS2 & Grating & 0.04 & 0.7 & 5.9 & 2.9 & 12.6 & 100 & {\bf 34}\\
\hline
\xrism{} & Resolve & TES & 0.2 & 5.0 & 0.7 & 0.3 & 4 & 32 & {\bf 107} \\
\athena{} & X-IFU & TES & 0.075 & 1.6 & 0.02 & 0.026 & 10 & 79 & {\bf 3038} \\
\hubs{} & XQSC & TES & $<$0.03 & $<$0.5 & $<$0.7 & $<$0.3 & $>$15.8 & $>$125 & {\bf $>$417} \\
\hline 
\arcus{} & & Grating & 0.006 & 0.1 & 2 & 1 & 40 & 316 & {\bf 316} \\
\lynx{} & XGS & Grating & $<0.003$ & $<0.05$ & $<0.8$ & $<0.4$ & $>50$ & $>395$ & $>${\bf 988} \\
\hline
\end{tabular}
\noindent
\end{table}

Generally, dispersive spectrometers, with no intrinsic limitations on resolving power, score better than microcalorimeters in detectability-power (FoM$_{det}$: Table \ref{tab:future_spec}), and can perform LSS absorption studies down to lower ion column densities (Table \ref{tab:future_spec}, Col. 5). 
However, only missions that combine a large throughput with a high resolving power can guarantee high investigation-powers (FoM$_{sdt}$). 
Particularly the \athena{} observatory scores by far the highest in FoM$_{std}$. It will allow the study of about 3000 LSS absorbers in a full year of observations of bright targets down to a column density of \ovii{} N$_{OVII} \simeq 1.6\times 10^{15}$ cm$^{-2}$. This will enable the systematic investigation of the physics, kinematics and chemistry of the most massive LSS gaseous structures, permeating halos of thousands $\ge$L$^*$ galaxies with masses in the $10^{12}-10^{13.5}$ M$_{\odot}$ range \citep{Wijers20}.  
The \hubs{} mission, should allow to take these studies down to three times smaller \ovii{} column densities, corresponding to halo masses of $\simeq 10^{11.5}$, while high resolving-power grating missions like \arcus{} and \lynx{} will extend these systematic studies down to the halos of 0.05L$^*$ galaxies (e.g. \citealp{Wijers20}), so uncovering the entire spectrum of gaseous baryons and metals in the LSS. 

\vspace{0.3cm}

%\section{References}
\bibliography{Gas-LSS.bib}

\end{document}